\documentclass[graybox]{svmult}

\usepackage{mathptmx}       
\usepackage{helvet}         
\usepackage{courier}        
\usepackage{type1cm}        
\usepackage{amsmath,amssymb,bm}
\usepackage{makeidx}         
\usepackage{graphicx}        
\usepackage{multicol}        
\usepackage[bottom]{footmisc}

\makeindex             

\newcommand{\MU}{{m}}
\newcommand{\rad}{{\text{rad}}}
\newcommand{\ret}{{\text{ret}}}
\newcommand{\adv}{{\text{adv}}}
\newcommand{\act}{{\text{act}}}
\newcommand{\new}{{\text{new}}}
\newcommand{\old}{{\text{old}}}
\newcommand{\con}{{\text{con}}}
\newcommand{\dis}{{\text{dis}}}
\newcommand{\schw}{{\text{Schw}}}
\newcommand{\s}{{\text{S}}}
\newcommand{\R}{{\text{R}}}

\newcommand{\calB}{{\mathcal{B}}}
\newcommand{\calE}{{\mathcal{E}}}
\newcommand{\calF}{{\mathcal{F}}}
\newcommand{\calR}{{\mathcal{R}}}

\newcommand{\Deqn}[1]{{Eq.~(\ref{#1})}}
\newcommand{\Deqns}[1]{{Eqs.~(\ref{#1})}}

\newcommand{\beq}{\begin{equation}}
\newcommand{\eeq}{\end{equation}}

\newcommand{\bea}{\begin{eqnarray}}
\newcommand{\eea}{\end{eqnarray}}

\begin{document}

\title*{Elementary development of the gravitational self-force}
\titlerunning{Elementary development of the gravitational self-force}
\author{Steven Detweiler}

 \institute{Steven Detweiler \at Institute of Fundamental Theory,
 Department of Physics, University of Florida, Gainesville,
 Florida, \email{det AT ufl.edu}}


\maketitle

 \abstract*{ The gravitational field of a particle of small mass $\MU$
 moving through curved spacetime, with metric $g_{ab}$, is naturally and
 easily decomposed into two parts each of which satisfies the perturbed
 Einstein equations through ${\text{O}}(\MU)$. One part is an inhomogeneous field
 $h^\text{S}_{ab}$ which, near the particle, looks like the Coulomb $\MU/r$ field
 with tidal distortion from the local Riemann tensor. This singular field is
 defined in a neighborhood of the small particle and does not depend
 upon boundary conditions or upon the behavior of the source in either the
 past or the future. The other part is a homogeneous field $h^\text{R}_{ab}$. In
 a perturbative analysis, the motion of the particle is then best described
 as being a geodesic in the metric $g_{ab}+h^\text{R}_{ab}$. This geodesic motion
 includes all of the effects which might be called radiation
 reaction and conservative effects as well. }

 \abstract{ The gravitational field of a particle of small mass $\MU$
 moving through curved spacetime, with metric $g_{ab}$, is naturally and
 easily decomposed into two parts each of which satisfies the perturbed
 Einstein equations through ${\text{O}}(\MU)$. One part is an inhomogeneous field
 $h^\text{S}_{ab}$ which, near the particle, looks like the Coulomb $\MU/r$ field
 with tidal distortion from the local Riemann tensor. This singular field is
 defined in a neighborhood of the small particle and does not depend
 upon boundary conditions or upon the behavior of the source in either the
 past or the future. The other part is a homogeneous field $h^\text{R}_{ab}$. In
 a perturbative analysis, the motion of the particle is then best described
 as being a geodesic in the metric $g_{ab}+h^\text{R}_{ab}$. This geodesic motion
 includes all of the effects which might be called radiation
 reaction and conservative effects as well. }

\section{Introduction}
Newton's apple hangs in a tree. The force of gravity is balanced by
the force from a branch, and the apple is at rest. Later, the apple
falls and accelerates downward until it hits the ground.

Einstein's insight elevates the lowly force of gravity to exalted
status as a servant
of geometry. Einstein's apple, being sentient and hanging in a tree,
explains its own non-geodesic, non free-fall, accelerated motion as
being caused by the force it feels from the branch. When the apple is
released by the branch, its subsequent free-fall motion is geodesic
and not accelerated. The apple is freed from all forces and does not
accelerate until it hits the ground.

These two perspectives have differing explanations and differing
descriptions of the motion, but the actual paths through the events of
spacetime are the same.
 Newton's understanding that the gravitational mass is identical to the
inertial mass implies that a small object in free-fall moves along a
trajectory which is independent of the object's mass.
 Einstein's Equivalence Principle requires that a small object in
free-fall moves along a geodesic of spacetime, a trajectory which is
independent of the object's mass. Newton's free-fall motion and Einstein's
geodesic motion describe a small object as moving along one and the same
sequence of events in spacetime.

Thorne and Hartle \cite{ThorneHartle85} give a clear and careful
description of the motion of a small nearly-Newtonian object through
the geometry of spacetime. They conclude that such motion might have
a small acceleration, consistent with Newtonian analysis\footnote{If
the acceleration of gravity $\vec g$ differs significantly across a
large object, then the center of mass moves responding to some
average, over the object, of $\vec g$ which does not necessarily
match a free-fall trajectory.}, from the coupling of the internal
mass multipole moments of the object with the multipole moments of
the external spacetime geometry, which are related to the components
of the Riemann tensor in the vicinity of the object [\textit{cf.}
Eqs.~(\ref{Eij})--(\ref{Bijk})]. If the object orbits a large black
hole, then the analysis implies that the motion is geodesic as long
as any asphericity of the object, perhaps caused by rotation or tidal
distortion, can be ignored. An acceleration larger than allowed by
the coupling of the multipole moments is inexplicable in the context
of either General Relativity or of Newtonian gravity and must
necessarily result from some non-gravitational force.

How does the Thorne-Hartle description meld with the notion that
Einstein's apple orbits a black hole, emits gravitational waves,
radiates away energy and angular momentum, and cannot then move along a
geodesic of the black hole geometry? Radiation reaction is not a
consequence of any asphericity of the apple. Does the apple move along
a geodesic? Would the apple, being sentient, describe its own motion as
free-fall?

For the moment consider the familiar electromagnetic
radiation-reaction force on an accelerating charge $q$ as given below
in \Deqn{ALDforce}. A notable feature is that the force is
proportional to $q^2$. Consequently this force is often described as
resulting from the charge $q$ interacting with its own
electromagnetic field, and the force is called the electromagnetic
\textit{self-force}.

Similar language is used with gravitation, but in that case the force is
proportional to $m^2$ and the resulting acceleration is proportional to
$m$. In general terms, the gravitational self-force is said to be
responsible for any aspect of motion that is proportional to the mass $m$
of the object at hand.
 Yet, with either Newtonian gravity or General Relativity, the motion
of an object of small mass $m$ is independent of $m$.
\textit{Gravitational self-force} appears to be an oxymoron.

But, even Newtonian gravity contains a gravitational self-force. One
might describe the motion of the Moon about the Earth as free-fall in
the Earth's gravitational field and conclude that
\begin{eqnarray}
  m a = m \Big(\frac{2\pi}{T}\Big)^2 r &=& \frac{GMm}{r^2}
\end{eqnarray}
where $r$ is the radius of the Moon's orbit, so that the orbital period is
\begin{equation}
   T = \sqrt{4\pi^2r^3/GM}
\end{equation}

A more accurate description of the motion includes the influence of the
Moon back on the Earth. Then the Moon is in free-fall in the Earth's
gravitational field while the Earth orbits their common center of mass.
And the conclusion becomes
\begin{eqnarray}
 m \Big(\frac{2\pi}{T}\Big)^2 r &=& \frac{GMm}{r^2(1+m/M)^2}
\nonumber\\
   T &=& \sqrt{4\pi^2r^3/GM}\,[1+m/M] .
\label{newtonSF}
\end{eqnarray}

The mass of the Moon has an influence on its own motion in
\Deqn{newtonSF}, and this influence could be (although it rarely is)
described as a consequence of the Newtonian gravitational self-force.
Nevertheless, Newton's law of gravity still implies that the Moon does
not exert a net gravitational force on itself. The acceleration of the
Moon is still properly lined up with the gradient of the Earth's
gravitational potential, and the Moon's motion is described as
free-fall or geodesic, depending upon whether one is Newton or
Einstein.

To me it seems inappropriate to describe the presence of the $m/M$ term in
\Deqn{newtonSF} as resulting from the interaction of the Moon with it's
own gravitational field. Rather, the $m/M$ term arises because the Earth
orbits the common center of mass of the Earth-Moon system.

The conundrum of radiation reaction as being consistent with geodesic
motion can now be resolved. Einstein's apple orbiting a black hole must
move along a geodesic, but the geometry through which it moves is the
black hole metric disturbed by the presence of the apple. Nevertheless,
this disturbed metric is a vacuum solution of the Einstein equations in
the neighborhood of the apple. If the motion were not geodesic, then
the apple could not explain its own motion as being free-fall in a
vacuum gravitational field. Such motion would violate Newton's laws as
well as Einstein's Equivalence Principle.

Throughout this manuscript we focus on the self-force acting on small
objects which are otherwise in unconstrained, free-fall motion---this
includes the most interesting case of the two body problem in general
relativity. This specifically excludes forced motion from, for example,
a mass bouncing on the end of a spring. This restricted interest allows
us in a general way to avoid many mathematical complications of Green's
functions in curved spacetime and to rely instead on a strongly
intuitive perspective which may be backed up with detailed analysis.

The Newtonian self-force problem in this Introduction is expanded upon
in Sect.~\ref{newtonian}, where it becomes clear that careful
definitions of coordinates are difficult to come by, and that physics
is best described in terms of precisely defined and physically measurable quantities.


\subsection*{Outline}

In Sect.~\ref{E-M} \,we describe Dirac's \cite{Dirac38} classical
treatment of radiation reaction in the context of electricity and
magnetism in a language which mimics our approach to the gravitational
self-force and to an illustrative toy problem in
Sect.~\ref{toyproblem}.

Perturbation theory in General Relativity is described in
Sect.~\ref{pert0}, applied to locally inertial coordinates in
Sect.~\ref{pert1}, applied to a neighborhood around a point mass in
Sect.~\ref{pert2}, and used to describe a small object moving through
spacetime in Sect.~\ref{pert3}

The gravitational self-force is described in Sect.~\ref{gsf1stOrder},
which includes discussions of the conservative and dissipative effects
and of some different possible implementations of self-force analyses.

The important and yet very confusing issue of gauge freedom in
perturbation theory is raised in Sect.~\ref{gauge}. And an example of
gauge confusion in action is given in Sect.~\ref{gaugeconfusion}.

An outline of the necessary steps in a self-force calculation is
given in Sect.~\ref{steps}, and some recent examples of actual
gravitational self-force results are in Sect.~\ref{examples} and
\ref{deluT}. Sect.~\ref{fieldreg} describes a possible future approach
to self-force calculations which is amenable to a 3+1 numerical
implementation in the style of numerical relativity.

Concluding remarks are in Sect.~\ref{conclusion}.

\subsection*{Notation}

The notation matches that used in an earlier review by the
author\cite{Det05} and is described here and again later in context.

Spacetime tensor indices are taken from the first third of the alphabet
$a,b,\ldots,h$\;, indices which are purely spatial in character are taken
from the middle third, $i,j,\ldots,q$\; and indices from the last third
$r,s\ldots,z$\; are associated with particular coordinate components. The
operator $\nabla_a$ is the covariant derivative operator compatible with
the metric at hand. We often use $x^i = (x,y,z)$ for the spatial
coordinates, and $t$ for a timelike coordinate.
An overdot, as in $\dot\calE_{ij}$, denotes a time derivative along
a timelike worldline.
 The tensor $\eta_{ab}$ is the flat Minkowskii metric $(-1,1,1,1)$,
down the diagonal.
 The tensor $f_{kl}$ is the flat, spatial Cartesian metric $(0,1,1,1)$,
down the diagonal.
 The projection operator onto the two dimensional surface of a constant $r$
two sphere is $\sigma_{i}{}^{j} = f_{i}{}^{j} - x_i x^j/r^2$.
 A capitalized index, $A$, $B$, \dots emphasizes that the index is spatial
and tangent to such a two sphere. Thus when written as $\sigma_{AB}$ the
projection operator is exhibiting its alternative role as the metric of
the two-sphere.
 The tensor $\epsilon_{ijk}$ is the spatial Levi-Civita tensor, which
takes on values of $\pm1$ depending upon whether the permutation of the
indices are even or odd in comparison to $x,y,z$.
 A representative length scale $\calR$ of the geometry in the region of
interest in spacetime is the smallest of the radius of curvature, the
scale of inhomogeneities, and the time scale for changes along a
geodesic. Typically, if the region of interest is a distance $r$ away
from a massive object $M$, then $\calR^{-2}\sim M/r^3$ provides a
measure of tidal effects, and $\calR\sim$ an orbital period.

\section{Newtonian examples of self-force and gauge issues}
\label{newtonian}

Newtonian gravity self-force effects appeared in the introduction.
Why don't we discuss these effects in undergraduate classical
mechanics? The primary reason is that the Newtonian two-body problem
can be solved easily and analytically without mention of the
self-force. But in addition, a description of the Newtonian
self-force introduces substantial, unavoidable ambiguities which are
similar to the relativistic choice of gauge. Only because gauge
confusion haunts all of perturbation theory in General Relativity do
we now examine the Newtonian self-force using an elementary example
made unavoidably confusing.

Consider a smaller mass $m_1$ and a larger mass $m_2$ in circular
orbits of radii $r_1$ and $r_2$ about their common center of mass, so
\begin{equation}
  m_1 r_1 = m_2 r_2 \;.
\end{equation}
And their separation is
\begin{equation}
  R=r_1+r_2 = r_1(1+m_1/m_2) \;.
\end{equation}
Newton's law of gravity gives
\begin{equation}
 \frac{m_1 v_1^2}{r_1} = \frac{G m_1 m_2}{(r_1+r_2)^2} \;.
\end{equation}

The velocity $v_1$ of the small object could be measurable by a
redshift experiment. For this Newtonian system
\begin{eqnarray}
  v_1^2 &=& \frac{G m_2r_1}{(r_1+r_2)^2}
\nonumber\\
         &=& \frac{G m_2}{r_1(1+m_1/m_2)^2}\;,
\nonumber\\
         &=& \frac{G m_2}{r_1}(1-2m_1/m_2 + \ldots)\;.
\end{eqnarray}
Thus we could state that in the limit that $m_1\rightarrow 0$, the
gravitational self-force decreases the orbital speed $v_1$ by a
fractional amount $-m_1/m_2$. But, as an alternative, it is also true
that
\begin{eqnarray}
  v_1^2 &=& \frac{G m_2}{R(1+m_1/m_2)}
\nonumber\\
   &=& \frac{G m_2}{R}(1-m_1/m_2+\ldots) \;.
\end{eqnarray}
Thus we could equally well state that in the limit that
$m_1\rightarrow 0$, the gravitational self-force decreases the
orbital speed $v_1$ by a fractional amount $-m_1/2m_2$. Which would
be correct?

How does the ambiguity arise? In the first treatment, near by the
orbit the radius $r_1$ was implicitly held fixed while we took the
limit $m_1\rightarrow0$, and in that limit $R$ approaches $r_1$ from
above. In the second treatment the separation $R$ was implicitly held
fixed in the limit, and in that case $r_1$ approaches $R$ from below.
Which of these is the ``correct'' way to take the limit? When viewed
near by, which is a better description of the size of the orbit $r_1$
or $R$?

In this Newtonian situation there might be some specific reason to
make one choice rather than the other and the confusion could be
resolved by including the detail of which quantity is being held
fixed during the limiting process. But, in General Relativity for a
small mass $m_1$ orbiting a much more massive black hole $m_2$ the
ambiguity persists. After including self-force effects on the motion
of $m_1$, it would be tempting to state that the Schwarzschild
coordinate $r$ of $m_1$'s location should be held fixed while
$m_1\rightarrow0$ to reveal the true consequences of the
gravitational self-force.
  However, only the spherical symmetry of the exact Schwarzschild
geometry allows for the unambiguous definition of $r$.
  Whereas the actual perturbed geometry is not spherically symmetric
and has no natural $r$ coordinate.

A clear statement of a perturbative gauge choice (\textit{cf}
Sect.~\ref{gauge}) that fixes the gauge freedom 
can provide a mathematically well-defined quantity $r$ on the manifold.
But physics has no preferred gauge and has no preferred choice for $r$,
just as neither $r_1$ nor $R$ is preferred in this Newtonian example.


Rather than arguing the benefits of one gauge choice over another, it
is far better to discard the focus on the radius $r_1$ or the
separation $R$ of the orbit, and to consider only quantities that could
be determined with clear, unambiguous physical measurements. The
orbital frequency $\Omega$ could be determined from the periodicity of
the system, and the speed of the less massive component $v_1$ could be
measured via a Doppler shift. We now look for a relationship between
these two physically measurable quantities.

{} From the Newtonian analysis above,
\begin{equation}
  \Omega^2 = \frac{G m_2}{r_1(r_1+r_2)^2}
         = \frac{G m_2}{r_1^3(1+m_1/m_2)^2}
\end{equation}
so that
\begin{equation}
  r_1 = \Big[\frac{G m_2}{(1+m_1/m_2)^2}\Big]^{1/3} \Omega^{-2/3}
\end{equation}
and
\begin{eqnarray}
  v_1^2 &=& \frac{G m_2 r_1}{(r_1+r_2)^2}
\nonumber\\
      &=& \frac{G m_2}{r_1 (1+m_1/m_2)^2}
\nonumber\\
      &=& \frac{(G m_2\Omega)^{2/3}}{(1+m_1/m_2)^{4/3}} =
      (Gm_2\Omega)^{2/3}\Big(1-\frac{4}{3}\frac{m_1}{m_2} + \ldots\Big)
\end{eqnarray}
Next, it seems appropriate to define a quantity with units of length
in terms of the physically measurable $\Omega$,
\begin{equation}
 R_\Omega^3 = Gm_2/\Omega^2 .
\end{equation}
Now the velocity $v_1$ of the orbit and the orbital frequency
$\Omega$ are related by
\begin{eqnarray}
  v_1^2 &=& \frac{Gm_2}{R_\Omega}\Big(1-\frac{4}{3}\frac{m_1}{m_2} + \ldots\Big),
\end{eqnarray}
and in terms of these measurable quantities it is unambiguous to
state that the gravitational self-force changes $v_1$, for a fixed
$\Omega$ by a fractional amount $-2m_1/3m_2$.

This describes the effect of the self-force on two physically
measurable observables and thus qualifies as a true, unambiguous
self-force effect.

\section{Classical electromagnetic self-force}
\label{E-M}

The standard expression \cite{Jackson} for the electromagnetic
radiation reaction force on a charged particle $q$ is
\begin{equation}
 {\bf \calF}_\rad = \frac{2}{3} \frac{q^2}{c^3} \ddot{\mathbf{v}} .
\label{ALDforce}
\end{equation}
Equation (\ref{ALDforce}) has issues of interpretation, but it does
indeed describe the radiation reaction force when applied with care.

Dirac's \cite{Dirac38} derivation of \Deqn{ALDforce} is my favorite and
can be described in a way that blends rather well with my preferred
description of the self-force and the toy problem described in the next
section.

First, Dirac considers the causally interesting retarded
electromagnetic field $F^\ret_{ab}$ of an accelerating charge. But, he
also considers the advanced field $F^\adv_{ab}$ and then describes what
I call the electromagnetic \textit{singular source} $\text{S}$ field in
flat spacetime
\begin{equation}
 F^\text{S}_{ab} = \frac12(F^\ret_{ab}+F^\adv_{ab}) .
\label{tempFS}
\end{equation}
The field $F^\text{S}_{ab}$ might also be called the \textit{symmetric}
field, as in ``symmetric under reversal of causal structure.''
$F^\text{S}_{ab}$ has unphysical causal features, but it is an exact
solution to Maxwell's equations with a source. In curved spacetime the
definition of the
singular source S field is more complicated than in the flat-space version
of \Deqn{tempFS}.

Dirac next allows the charge $q$ to be of finite size. Then he presents
a subtle analysis using the conservation of the electromagnetic
stress-energy tensor in a neighborhood of the charge to show that
$F^\text{S} _{ab}$ exerts no net force on the charge in the limit that
the size of the charge is vanishingly small.

Now let $F^\act_{ab}$ be the actual, measurable electromagnetic field.
Then $F^\act_{ab}$ may be separated into two parts
\begin{eqnarray}
  F^\act_{ab} &=& F^\text{S}_{ab} + F^\text{R}_{ab}
\end{eqnarray}
where the remainder R-field is \textit{defined} by
\begin{eqnarray}
  F^\text{R}_{ab} &\equiv& F^\act_{ab} - F^\text{S}_{ab} .
\label{Rdef}
\end{eqnarray}
Both $F^\act_{ab}$ and $F^\text{S}_{ab}$ are solutions to Maxwell's
equations, in the neighborhood of $q$, with identical sources. Thus
$F^\text{R}_{ab}$ is necessarily a vacuum solution of the
electromagnetic field equations and is therefore regular in the
neighborhood of the particle.

Dirac then states that the radiation reaction force on the charge $q$
moving with four-velocity $u^a$ is
\begin{equation}
  {\bf\calF}_b^\rad = q u^a F^\text{R}_{ab}
\label{F-Rforce}
\end{equation}
and later shows that this is consistent with \Deqn{ALDforce}. In this
context $F^\text{R}_{ab}$ might be called the \textit{radiation
reaction} field, in view of the force it exerts on the charge.

Imagine the situation as viewed by a local observer who moves with the
particle and is able to measure and analyze the actual electromagnetic
field only in a neighborhood which includes the particle but is
substantially smaller than the wavelength of any radiation. The
observer is therefore not privy to any information whatsoever about
distant boundary conditions, or about the possible existence of
electromagnetically active material outside the neighborhood or even
about the possibility of electromagnetic radiation either ingoing or
outgoing at a great distance.

After considering the motion of the charge, the observer could
calculate $F^\text{S}_{ab}$ and then subtract it from the measured
$F^\act_{ab}$ to yield $F^\text{R}_{ab}$. Finally the observer could
apply
\Deqn{F-Rforce} and conclude that the Lorentz force law correctly
describes the electromagnetic contribution to the acceleration of the
charge, even though the observer might be completely unaware of the
presence of the radiation.

Thus $F^\act_{ab}$ is decomposed into two parts \cite{DetWhiting03}.
One part $F^\text{S}_{ab}$ is singular at the point charge, can be
identified as the particle's own electromagnetic field, and exerts no
force on the particle itself. The other part $F^\text{R}_{ab}$ does
exert a force on the particle, is a locally source-free solution of
Maxwell's equations and can be locally identified only as an externally
generated field of indeterminate origin. A local observer would have no
direct information about the source of $F^\text{R}_{ab}$ and, in
particular, could not distinguish the effects of radiation reaction
from the effects of boundary conditions.

\section{A toy problem with two length scales that creates a challenge for
numerical analysis} \label{toyproblem}

Binary inspiral of a small black hole into a much larger one presents
substantial difficulties to the numerical relativity community.
Perhaps the primary difficulty results from having two very different
length scales. On the one hand, a very coarse grid size would allow
easy resolution of the metric of the large black hole as well as
coverage out to the wavezone resulting in the efficient production of
gravitational waveforms. On the other hand, a very fine grid size
would provide the detailed information about the metric in a
neighborhood of the small black hole necessary for tracking the
evolution of the binary system and for providing accurate
gravitational waveforms.

The following toy problem shares the two length-scale difficulty of
binary inspiral. But it is elementary, not complicated by curved
spacetime or subtle dynamics, and yet leads to some insight on how the
binary inspiral problem might be approached. In addition, its
resolution involves some aspects of Dirac's analysis of electromagnetic
radiation reaction as presented in the previous section.

Consider this flat space numerical analysis problem in electrostatics:
An object of small radius $r_\text{o}$ has a spherically symmetric
electric charge density $\rho(r)$ with an associated electrostatic
potential $\varphi$. The object is inside an odd shaped grounded,
conducting box which is much larger than $r_\text{o}$. The boundary
condition on the potential is that $\varphi=0$ on the box. For
simplicity assume that the small object is at rest at the origin of
coordinates. Thus, there is no radiation and the field equation for
$\varphi$ is elliptic. Then
\begin{equation}
  \nabla^2\varphi = -4\pi \rho
\label{del2psi}
\end{equation}
where $\vec\nabla$ is the usual three-dimensional flat space gradient
operator, and $\nabla^2$ the Laplace operator. Let $\vec r$ refer to
the displacement from the center of the object at the origin to a
general point in the domain within the box.

Here is the goal: Given $\rho(r)$, numerically determine $\varphi$ as
a function of $\vec r$ everywhere inside the box, subject to the
field equation (\ref{del2psi}) and to the boundary condition that
$\varphi=0$ on the boundary of the box. Then find the total force on
the small object which results from its interaction with $\varphi$.

Here is the difficulty: If the object is much smaller than the box,
then the difference in length scales complicates calculating $\varphi$.
The object is very small so an accurate analysis would require a very
fine grid size. However, the distance from the object to the boundary
of the box is large compared to the size of object. Thus a relatively
coarse grid size would be desired to speed up the numerical evaluation.
The difficulty is exacerbated if we are also interested in the force
from $\varphi$ acting back on the object; this requires accurately
knowing the value of $\varphi$ inside the small object precisely where
$\varphi$ has substantial variability.

We will shortly introduce a variety of versions of the potential under
consideration. For clarity, the \textit{actual} electrostatic potential
$\varphi^\act$ actually satisfies both the field equation
(\ref{del2psi}) with the actual source and also the relevant boundary
conditions. Thus, $\varphi^\act$ is the potential which an observer
would actually measure for the problem at hand.

\subsection{An approach which avoids the small length scale}

To remove the two-length-scale numerical difficulty we take the
following approach: In a neighborhood of the object the potential ought
to be approximated by the function $\varphi^\text{S}$ defined as the
usual electrostatic potential of a spherical distribution of charge
which for a constant charge density $\rho(r)$ and total charge $q$ is
\begin{eqnarray}
 \text{for} \quad r< r_\text{o}:\quad \varphi^\text{S}(r) &=&
                                 \frac{q}{2r_\text{o}^3}(3r_\text{o}^2-r^2)
\nonumber\\
 \text{for} \quad r> r_\text{o}:\quad \varphi^\text{S}(r) &=& q/r .
\label{defpsiS}
\end{eqnarray}
The \textit{source} field $\varphi^\text{S}(r)$ is completely
determined by local considerations in the neighborhood of the object,
and it is chosen carefully to be an elementary solution of
\begin{equation}
  \nabla^2 \varphi^\text{S}= -4\pi \rho .
\label{delpsiS}
\end{equation}
Sometimes $\varphi^\text{S}$ is called the \textit{singular} field to
emphasize the $q/r$ behavior outside but near a small source.
 Viewed from near by, the actual field $\varphi^\act$ is approximately
$\varphi^\text{S}$.

Given $\varphi^\text{S}$, the numerical problem may be reformulated in
terms of the field
\begin{equation}
  \varphi^\text{R} \equiv \varphi^\act - \varphi^\text{S}
  \label{defpsiR}
\end{equation}
which is then a solution of
\begin{eqnarray}
  \nabla^2 \varphi^\text{R} &=& - \nabla^2\varphi^\text{S}- 4 \pi \rho = 0,
  \label{del2psiR}
\end{eqnarray}
where the second equality follows from \Deqn{delpsiS}. The
\textit{regular} field $\varphi^\text{R}$ is thus a source free
solution of the field equation, and is sometimes called the
\textit{remainder} when the \textit{subtrahend} $\varphi^\text{S}$ is
removed from the actual field $\varphi^\act$ in Eq.~(\ref{defpsiR}).

Viewed from afar, the boundary condition that $\varphi^\act = 0$ on the
box plays an important role and determines the boundary condition that
$\varphi^\text{R} = - \varphi^\text{S}$ on the box. Thus, rewriting the
problem in terms of the analytically known $\varphi^\text{S}$ and the
``to be determined numerically'' $\varphi^\text{R}$ leaves us with the
boundary value problem
\begin{equation}
  \nabla^2 \varphi^\text{R} = 0\quad\text{ with the boundary condition that
  $\varphi^\text{R} = -\varphi^\text{S}$ on the box. }
\label{del2psiRA}
\end{equation}
It is important to note that $\varphi^\text{R}$ is a regular,
source-free solution of the field equation.

In this formulation based upon \Deqn{del2psiRA} $\varphi^\text{R}$
scales as the charge $q$ but has no structure with the length scale of
the source $r_\text{o}$. The small length scale has been completely
removed from the problem. The removal is at the expense of introducing
a complicated boundary condition---but at least the boundary condition
does not have an associated small length scale. Once $\varphi^\text{R}$
has been determined, the actual field $\varphi^\act =
\varphi^\text{R}+\varphi^\text{S}$ is easily constructed.\footnote{Following
Dirac's\cite{Dirac38} usage, I prefer to use the word ``actual'' to
refer to the complete, and total field that might be measured at some
location. Often in self-force treatises the ``retarded field'' plays
this central role. But, this obscures the fact that, viewed from near
by, a local observer unaware of boundary conditions could make no
measurement which would reveal just what part of the field is the
retarded field. This confusion is increased if the spacetime is not
flat, so that the retarded field could be determined only if the entire
spacetime geometry were known.}

But that's not all: This formulation has the bonus that it simplifies
the calculation of the force on the object from the field. The force is
an integral over the volume of the object,
\begin{equation} 
  \vec F = -\int \rho(r) \vec\nabla\varphi^\act\, d^3x.
\label{selfF}
\end{equation}
In the original formulation using \Deqn{del2psi}, the actual field
$\varphi^\act$ in the integral would be dominated by $\varphi^\text{S}$
which changes dramatically over the length scale of the object, and
$\varphi^\text{R}$ could be easily lost in the noise of the
computation. The spherical symmetry of $\varphi^\text{S}$ and $\rho$
imply that
\begin{eqnarray}
 \int \rho(r) \vec\nabla\varphi^\text{S}\, d^3x = 0.
\label{intrhopsiS}
\end{eqnarray}
Then the substitution $\varphi^\act\rightarrow\varphi^\text{S}+
\varphi^\text{R}$ in the integral of \Deqn{selfF} leads to the conclusion
that
\begin{equation}
  \vec F = -\int \rho(r) \vec\nabla\varphi^\text{R}\, d^3x .
\label{F_R}
\end{equation}
Thus the force acting on the object may be written in terms of only
$\varphi^\text{R}$.

But that's not all: The field $\varphi^\text{R}$ does not change
significantly over a small length scale, so if the object is extremely
small (Think: an approximation to a $\delta$-function.) then an
accurate approximation to the force is
\begin{equation}
  \vec F = -q \vec\nabla\varphi^\text{R}|_{r=0},
\label{F-R}
\end{equation}
when viewed from near by.

Standard jargon calls the force in \Deqn{F-R} the ``self-force''
because it is necessarily proportional to $q^2$ and apparently
results from the object interacting with ``its own field.'' But, it
is important to note that this force clearly depends upon the shape
of the box, \textit{i.e.} the details of the boundary conditions. In
my opinion the physics appears more intuitive to have ``the object's
own field,'' refer only to $\varphi^\text{S}$ whose local behavior is
defined uniquely and independently of any boundary conditions. And
$\varphi^\text{S}$ is also guaranteed to exert no force back on the
charge. Then $\varphi^\text{R}$ is a regular source-free solution to
the field equation in the neighborhood of the object and is solely
responsible for the force acting on the object. An observer local to
the object would know $\rho(r)$, could calculate $\varphi^\text{S}$
and measure $\varphi^\act$.
  Subtracting $\varphi^\text{S}$ from the actual field $\varphi^\act$
then results in the regular remainder $\varphi^\text{R}=\varphi^\act -
\varphi^\text{S}$. While the force described in Eq.~(\ref{F-R}) is
indeed proportional to $q^2$, it still seems sensible to refer to this
as simply ``the force'' on the object.

\subsection{An alternative that resolves boundary condition issues}
 \label{alternative}

The previous resolution of the difficulty of the two length scales
caused a change and complication of the boundary conditions. With a
slight variation, the problem can be reformulated in a way that
brings back the original, natural boundary conditions.

The alternative approach deals with the boundary condition
complication by introducing a window function $W(r)$
\cite{VegaDet08} which has three properties:
\begin{enumerate}
\item[A.] $W(r) = 1$ in a region which includes at least the
    entire source $\rho(r)$, that is all $r \le r_\text{o}$.

\item[B.] $W(r)=0$ for $r > r_{\text{W}}$ where $r_{\text{W}}$ is
    generally much larger than $r_\text{o}$ but is restricted so
    that the entire region $r<r_\text{W}$ is inside the box.

\item[C.] $W(r)$ is $C^\infty$ and changes only over a long
    length scale comparable to $r_{\text{W}}$.
\end{enumerate}

For this alternative approach the field defined by
\begin{equation}
  \varPhi^\text{R} \equiv \varphi^\act- W\varphi^\text{S}
  \label{defpsiRP}
\end{equation}
is a solution of
\begin{eqnarray}
  \nabla^2 \varPhi^\text{R} &=& - \nabla^2(W \varphi^\text{S}) - 4 \pi \rho
\nonumber\\ &=&
    - \varphi^\text{S}\nabla^2W - 2\vec\nabla W\cdot \vec\nabla \varphi^\text{S}
             - W\nabla^2\varPhi^\text{S}- 4 \pi \rho
\nonumber\\ &=&
    - \varphi^\text{S}\nabla^2W - 2\vec\nabla W\cdot \vec\nabla \varphi^\text{S}
    \equiv S_\text{eff},
\label{del2psiRPa}
\end{eqnarray}
where $S_\text{eff}$ is the \textit{effective source} and the third
equality follows from \Deqn{delpsiS} and property (A). The boundary
condition is now that $\varPhi^\text{R} = 0$ on the box, which is the
natural boundary condition. Thus, rewriting the problem in terms of the
analytically known $\varphi^\text{S}$ and the to-be-determined-numerically
$\varPhi^\text{R}$ leaves us with the field equation
\begin{equation}
  \nabla^2 \varPhi^\text{R} = S_\text{eff}
\label{del2psiRP}
\end{equation}
and the \textit{natural} boundary condition that $\varPhi^\text{R} = 0$
on the box.

It is important to note that the effective source $S_\text{eff}$ defined
in \Deqn{del2psiRPa} is zero inside the small object where $W(r) = 1$ and
changes only over a long length scale $r_{\text{W}}$. Thus the field
$\varPhi^\text{R}$ is a regular, source-free solution of the field
equation inside the object, and outside the object $\varPhi^\text{R}$ only
changes over a long length scale $r_{\text{W}}$. And Eqs.~(\ref{F_R}) and
(\ref{F-R}) provide the force acting on the object, after
$\varphi^\text{R}$ is replaced with $\varPhi^\text{R}$.

This alternative approach completely removes the small length scale
from the problem and leaves the natural boundary condition
$\varPhi^\text{R} = 0$ on the box intact.

In applications of this approach to problems in curved spacetime, the
singular field $\varphi^\text{S}$ is rarely known exactly. In fact, for
a $\delta$-function source often only a finite number of terms in an
asymptotic expansion are available. This limits the differentiability
of the source of \Deqn{del2psiRP} which, in turn, limits the
differentiability of $\varPhi^\text{R}$ at the particle. But the
procedure remains quite adequate for solving self-force problems.

This approach to the self-force, which introduces a window function,
has now been implemented for a scalar charge in a circular orbit of the
Schwarzschild geometry and is discussed below in Sect.~\ref{fieldreg}.

\section{Perturbation theory}
\label{perturbations}
\label{gravSF}

Perturbation theory has had some great successes in General
Relativity particularly in the realm of black holes
\cite{ReggeWheeler, Zerilli, TeukolskyI, SasakiNakamura} by proving
stability \cite{ReggeWheeler, Zerilli, Vishveshwara, Whiting},
analyzing the quasi-normal modes, \cite{Press71, Quasi75, Det79,
Leaver85, Leaver86} and calculating the gravitational waves from
objects falling in and around black holes \cite{Zerilli, DRPP,
CunninghamMP78, CunninghamMP79, DetweilerIII} to highlight just a few
of the earlier accomplishments.

In preparation for the era of gravitational wave astronomy,
relativists are now turning their attention to second and higher
order perturbation analysis. However, we focus on linear order and
give a brief description of this theory.

In Sect.~\ref{pert0} we begin with an overview that emphasizes the
Bianchi identity's implication that a perturbing stress-energy tensor
$T_{ab}$ must be conserved $\nabla^aT_{ab}=0$ to have a well formulated
perturbation problem. This requires that an object of small size and
mass must move along a geodesic.

We use perturbation theory in Sect.~\ref{pert1} to describe the
geometry in the vicinity of a timelike geodesic $\Gamma$ of a vacuum
spacetime. We specifically use a locally inertial and harmonic
coordinate system, THZ coordinates introduced by Thorne and Hartle
\cite{ThorneHartle85}, to represent the metric as a perturbation of
flat spacetime $g_{ab} = \eta_{ab}+H_{ab}$ in a particularly
convenient manner within a neighborhood of the geodesic.

In Sect.~\ref{pert2} we put a small mass $\MU$ down on this same
geodesic $\Gamma$ and treat its gravitational field
$h^{\text{S}}_{ab}$ as a perturbation of $g_{ab}$.

Finally, in Sect.~\ref{pert3} we identify $h^{\text{S}}_{ab}$ as the
S-field of $\MU$, the analogue of $F^{\text{S}}_{ab}$ in Sect.~\ref{E-M}
and of $\phi^{\text{S}}$ in Sect.~\ref{toyproblem}. In particular
$h^{\text{S}}_{ab}$ is a metric perturbation which is singular at the
location of $\MU$, is a solution of the field equation for a
$\delta$-function point mass and exerts no force back on the mass $\MU$
itself.

\subsection{Standard perturbation theory in General Relativity}
\label{pert0}

We start with a spacetime metric $g_{ab}$ which is a vacuum solution
of the Einstein equations $G_{ab}(g)=0$. Then we ask, ``What is the
slight perturbation $h_{ab}$ of the metric created by a small object
moving through the spacetime along some worldline $\Gamma$?''

Let $\calR$ be a representative length scale of the geometry near the
object which is the smallest of the radius of curvature, the scale
of inhomogeneities, and the time scale for changes in curvature along
the world line of the object. When we say ``small object'' we imply
that the size $d$ of the object is much less than $\calR$ and that
the mass $\MU$ is much smaller than $d$.

As a notational convenience, the Einstein tensor $G_{ab}(g+h)$ for a
perturbed metric may be expanded in powers of $h$ as
\begin{equation}
  G(g+h) = G(g) + G^{(1)}(g,h) + G^{(2)}(g,h) + \ldots
\label{G0+G1+G2}
\end{equation}
where $G^{(n)}(g,h) = {\text{O}}(h^n)$. The zeroth order term $G(g)$
is zero if $g_{ab}$ is a vacuum solution of the Einstein equations. The
first order part is $G^{(1)}_{ab}(g,h)$, which resembles a linear wave
operator on $h_{ab}$ and is equivalent to the operator $-E_{ab}(h)$
given below in
\Deqn{Eab}.  
The second order part $ G^{(2)}(g,h)$
consists of terms such as ``$\nabla h \nabla h$'' or ``$ h \nabla
\nabla h$,'' similar to the Landau-Lifshitz pseudo tensor \cite{LandL}.
The third and higher order terms in the expansion (\ref{G0+G1+G2}) are
less familiar.

Next, we assume that the stress-energy tensor of the object $T_{ab}$
is ${\text{O}}(\MU)$, and that the perturbation in the metric
$h_{ab}$ is also ${\text{O}}(\MU)$. At first perturbative order,
\begin{equation}
  G_{ab}(g+h) = 8\pi T_{ab} + \text{O}(h^2).
\label{einsteinh}
\end{equation}

We expand $G_{ab}(g+h)$ through first order in $h$
via the symbolic operation
\begin{equation}
   G_{ab}^{(1)}(g,h) = \frac{\delta G_{ab}}{\delta g_{cd}} h_{cd}
\label{Gabdef}
\end{equation}
and define the wave operator mentioned above by $E_{ab}(h) \equiv - G_{ab}^{(1)}(g,h)$,
so that
\begin{eqnarray}
  2E_{ab}(h) &=& \nabla^2 h_{ab} + \nabla_a \nabla_b h
           - 2 \nabla_{(a}\nabla^c h_{b)c}
\nonumber\\ & &
      + 2{R_a}^c{}_b{}^d h_{cd} 
      + g_{ab} ( \nabla^c\nabla^d h_{cd} - \nabla^2 h ) ,
\label{Eab}
\end{eqnarray}
with $h \equiv h_{ab} g^{ab}$. Also $\nabla_a$ and ${{R_a}^c}_b{}^d$ are
the derivative operator and Riemann tensor of $g_{ab}$.
 If $h_{ab}$ solves
\begin{equation}
  E_{ab}(h) = - 8\pi T_{ab} .
\label{EabTab}
\end{equation}
then \Deqn{einsteinh} is satisfied.

In an actual project, the biggest technical task is usually solving
Eq.~(\ref{EabTab}). As an example, the study of gravitational
radiation from an object orbiting a Schwarzschild black hole
typically invokes the Regge-Wheeler-Zerilli formalism
\cite{ReggeWheeler, Zerilli}.

With a vacuum-spacetime metric $g_{ab}$ and \textit{any} symmetric
tensor $k_{ab}$, the Bianchi identity implies that
\begin{equation}
  \nabla^a E_{ab}(k) = 0.
\end{equation}
This is easily demonstrated by direct analysis, after starting with
\Deqn{Eab}. Thus, for a solution of \Deqn{EabTab} to exist, it is
necessary that the integrability condition
\begin{equation}
  \nabla^a T_{ab} = 0
\label{DaTab}
\end{equation}
for the stress-energy tensor be satisfied.

If the stress-energy tensor is only approximately conserved $\nabla^a
T_{ab} = {\text{O}}(\MU^2)$ then the solution for $h_{ab}$ might be in
error at $\text{O}(\MU^2)$. In some circumstances this might be
acceptable, in which case if $T_{ab}$ represents the stress-energy tensor
for a particle of small size, then the particle must move along an
approximate geodesic of $g_{ab}$ \cite{PoissonLR} with an acceleration no
larger than ${\text{O}}(\MU)$. Then the integrability condition is nearly
satisfied and $h_{ab}$ can be determined from \Deqn{EabTab}.

Next, one might be inclined to attempt the analysis of the Einstein
equations through second order in the perturbation $h_{ab}$. But,
this requires that $T_{ab}$ be conserved, not in the metric $g_{ab}$,
but rather in the first order perturbed metric $g_{ab}+h_{ab}$. Thus
the worldline of a particle is not geodesic in $g_{ab}$ and its
acceleration as measured in $g_{ab}$ is often said to result from the
\textit{gravitational self-force}. After the self-force problem is
solved for the $\text{O}(\MU)$ adjustment to the motion of the
particle, then the second order field equation from
Eq.~(\ref{G0+G1+G2}) determines $h_{ab}$ through ${\text{O}}(\MU^2)$.

As described by Thorne and Kov\'{a}cs \cite{ThorneKovacs75}, this
process continues: With the improved metric, solve the dynamical
equations for a more accurate worldline and stress-energy tensor.
With the improved stress-energy tensor solve the field equations for
a more accurate metric perturbation. Repeat.

This alternation of focus between the dynamical equations and the field
equations is quite similar to that used in post-Newtonian analyses.

\subsection{An application of perturbation theory: locally inertial coordinates}

\label{pert1}

Before dealing with perturbing masses, we first consider vacuum
perturbations of a vacuum spacetime and focus on a neighborhood of a
timelike geodesic $\Gamma$ where the metric appears as a perturbation
$H_{ab}$ of the flat Minkowskii metric $\eta_{ab}$.

This application is simplified by use of a convenient coordinate
system described by Thorne and Hartle \cite{ThorneHartle85}. It is
well known in General Relativity \cite{Weinberg}, that for a timelike
geodesic $\Gamma$ in spacetime there is a class of \textit{locally
inertial} coordinate systems $x^a=(t,x,y,z)$, with $r^2=x^2+y^2+z^2$,
which satisfies the following conditions:
\begin{enumerate}
\item[A.] The geodesic $\Gamma$ is identified with $x=y=z=r=0$
    and $t$ measures the proper time along the worldline.
\item[B.] On $\Gamma$, the metric takes the Minkowskii form
    $g_{ab} = \eta_{ab}$.
\item[C.] All first derivatives of $g_{ab}$ vanish on $\Gamma$ so
    that the Christoffel symbols also vanish on $\Gamma$.
\end{enumerate}
Fermi-normal coordinates \cite{MTW} provide an example which meets
all of these locally inertial criteria.

With a locally inertial coordinate system in hand, it is natural to Taylor
expand $g_{ab}$ about $\Gamma$ with
\begin{equation}
 g_{ab} = \eta_{ab} + H_{ab} + \ldots
\end{equation}
where
\begin{eqnarray}
  H_{ab} &=& {}_2H_{ab}+{}_3H_{ab} ,
\nonumber\\
  {}_2H_{ab} &=& \frac{1}{2} x^i x^j \partial_i\partial_j\, g_{ab} ,
\nonumber\\
  {}_3H_{ab} &=& \frac{1}{6} x^i x^j x^k \partial_i\partial_j\partial_k\,
  g_{ab} ,
\end{eqnarray}
and the partial derivatives are evaluated on $\Gamma$.

The quantities ${}_2H_{ab}$ and ${}_3H_{ab}$ scale as
${\text{O}}(r^2/\calR^2)$ and ${\text{O}}(r^3/\calR^3)$ in a small
neighborhood of $\Gamma$, and these may be treated as perturbations
of flat spacetime with $r/\calR$ being the small parameter. Recall
that $\calR$ is a length scale of the background geometry. First
order perturbation theory is applicable here because $H_{ab}$ has no
${\text{O}}(r/\calR)$ term but starts at ${\text{O}}(r^2/\calR^2)$.
Thus ${}_2H_{ab}$ and ${}_3H_{ab}$ may be treated as independent
perturbations and the first nonlinear term appears at
${\text{O}}(r^4/\calR^4)$. Thus, $H_{ab}$ is a perturbation which
must satisfy the source-free perturbed Einstein equations $E_{ab}(H)
= 0$.

Thorne and Hartle \cite{ThorneHartle85} and Zhang \cite{Zhang86} show
that a particular choice of locally inertial coordinates leads to a
relatively simple expansion of the metric. Initially they introduce
spatial, symmetric, trace-free \textit{multipole moments} of the
external spacetime $\calE_{ij}$, $\calB_{ij}$, $\calE_{ijk}$, and
$\calB_{ijk}$ which are functions only of $t$ and are directly
related to the Riemann tensor evaluated on $\Gamma$ by
\begin{equation}
 \calE_{ij} = R_{titj},
\label{Eij}
\end{equation}
\begin{equation}
  \calB_{ij} = \epsilon_i{}^{pq}R_{pqjt}/2,
\end{equation}
\begin{equation}
  \calE_{ijk} =\left[\partial_kR_{titj}\right]^{\mbox{\scriptsize STF}}
\end{equation}
and
\begin{equation}
   \calB_{ijk} = \frac{3}{8}
     \left[\epsilon_i{}^{pq}\partial_kR_{pqjt}
                  \right]^{\mbox{\scriptsize STF}}.
\label{Bijk}
\end{equation}
Here ${}^{\mbox{\scriptsize STF}}$ means to take the symmetric,
tracefree part with respect to the spatial indices, and
$\epsilon_{ijk}$ is the flat, spatial Levi-Civita tensor, which takes
on values of $\pm1$ depending upon whether the permutation of the
indices are even or odd in comparison to $x,y,z$.
 Also, $\calE_{ij}$ and $\calB_{ij}$ are ${\text{O}}(1/\calR^2)$, while
$\calE_{ijk}$ and $\calB_{ijk}$ are ${\text{O}}(1/\calR^3)$.
 All of the above multipole moments are tracefree because the external
background geometry is assumed to be a vacuum solution of the Einstein
equations.

Spatial STF tensors are closely related to linear combinations of
spherical harmonics. For example the STF tensor $\calE_{ij}$ with two
spatial indices is related to the $\ell=2$ spherical harmonics
$Y_{2,m}$ by
\begin{equation}
  \calE_{ij}x^ix^j = r^2 \sum_{m=-2}^2 E_{2,m} Y_{2,m},
\end{equation}
with the five independent components of $\calE_{ij}$ being determined
by the five independent coefficients $E_{2,m}$.

Next an infinitesimal coordinate transformation (a perturbative gauge
transformation, Sect.~\ref{gauge}) changes the description of
$H_{ab}$ to a form where the partial derivatives in the Taylor
expansion are equivalent to the components of the Riemann tensor and
represented by the multipole moments. The result is
\begin{eqnarray}
   {}_2H_{ab}d x^a d x^b & = &
         - \calE_{ij} x^i x^j ( {\,{\rm d}t}^2 + f_{kl} {{\rm d}} x^k {\,{\rm d}x}^l )
         {} + \frac{4}{3} \epsilon_{kpq}\calB^q{}_i x^p x^i {\,{\rm d}t} {\,{\rm d}x}^k
\nonumber\\ &&
         {} - \frac{20}{21} \Big[ \dot{\calE}_{ij}x^ix^j x_k
            - \frac{2}{5} r^2 \dot{\calE}_{ik} x^i \Big] {\,{\rm d}t} {\,{\rm d}x}^k
\nonumber\\ &&
 {} + \frac{5}{21} \Big[ x_i \epsilon_{jpq} \dot{\calB}^q{}_k x^px^k
    - \frac{1}{5} r^2 \epsilon_{pqi}
                 \dot{\calB}_{j}{}^q x^p \Big] \,{\,{\rm d}x}^i\, {\,{\rm d}x}^j
    + {\text{O}}( r^4/\calR^4)
\label{H2}
\end{eqnarray}
and
\begin{eqnarray}
  {}_3H_{ab}{{\rm d}} x^a {{\rm d}} x^b & = &
         - \frac13\calE_{ijk} x^i x^j x^k
            ( {\,{\rm d}t}^2 + f_{lm} {\,{\rm d}x}^l {\,{\rm d}x}^m )
\nonumber\\ &&
         {} + \frac{2}{3} \epsilon_{kpq}\calB^q{}_{ij}
             x^p x^i x^j {\,{\rm d}t} {\,{\rm d}x}^k
          + {\text{O}}( r^4/\calR^4),
\label{H3}
\end{eqnarray}
where $f_{kl}$ is the flat, spatial Cartesian metric $(0,1,1,1)$,
down the diagonal. The overdot represents a time derivative along
$\Gamma$ of, say, $\calE_{ij}={\text{O}}(\calR^{-2})$, and then $\dot
\calE_{ij} = {\text{O}}(\calR^{-3})$ because $\calR$ bounds the time
scale for variation along $\Gamma$.

A straightforward evaluation of the Riemann tensor for the metric
$\eta_{ab}+{}_2H_{ab}+{}_3H_{ab}$ confirms that the STF multipole
moments are related to the Riemann tensor as claimed in
\Deqns{Eij}--(\ref{Bijk}).

We call the locally inertial coordinates of Thorne, Hartle and Zhang
used in \Deqns{H2} and (\ref{H3}) THZ coordinates.

If interest is focused only on the lower orders ${\text{O}}(r^2/\calR^2)$
and ${\text{O}}(r^3/\calR^3)$, then THZ coordinates are not unique and
freedom is allowed in their construction away from the worldline $\Gamma$.
Given one set of THZ coordinates $x^a$, a new set defined from
$x^a_\text{new} = x^a + \lambda^a_{ijklm}x^ix^jx^kx^lx^m$, where
$\lambda^a_{ijklm}={\text{O}}(1/\calR^4)$ is an arbitrary function of
proper time on $\Gamma$, preserves the defining form of the expansion
given in \Deqns{H2} and (\ref{H3}).

Work in preparation 
describes a direct, constructive
procedure for finding a THZ coordinate system associated with any
geodesic of a vacuum solution of the Einstein equations.

\subsection{Metric perturbations in the neighborhood
   of a point mass.}
\label{pert2}

We are now prepared to use perturbation theory to determine
$h^{\text{S}}_{ab}$, the gravitational analogue of
$F^{\text{S}}_{ab}$ in Sect.~\ref{E-M} and of $\varphi^{\text{S}}$ in
Sect.~\ref{toyproblem}.


We consider the perturbative change $h_{ab}$ in the metric $g_{ab}$
caused by a point mass $\MU$ traveling through spacetime.
 We look for the solution $h_{ab}$ to \Deqn{EabTab} with the
stress-energy tensor $T^{ab}$ of a point mass
 \begin{equation}
  T^{ab} = \MU \int_{-\infty}^\infty \frac{ u^a u^b}{\sqrt{-g}}
       \delta^4(x^a-X^a(s)) \,\mbox{d}s
 \label{Tab}
 \end{equation}
where $X^a(s)$ describes the worldline of $\MU$ in an arbitrary
coordinate system as a function of the proper time $s$ along the
worldline.

The integrability condition for \Deqn{EabTab} requires the
conservation of $T^{ab}$, and we put $\MU$ down on the geodesic
$\Gamma$ of the previous section and limit interest to a neighborhood
of $\Gamma$ where $r^4/\calR^4$ is considered negligible although
$r^3/\calR^3$ is not.
And we use THZ coordinates. The perturbed metric of Sect.~\ref{pert1}
is now viewed as the ``background'' metric, $g_{ab} =
\eta_{ab}+H_{ab}$, with $H_{ab}$ given in Eqs.~(\ref{H2}) and
(\ref{H3}). The stress-energy tensor $T_{ab}$ for a point mass is
particularly simple in THZ coordinates and has only one nonzero
component
\begin{equation}
  T_{tt} = -\MU \delta^3(x^i) .
\label{Ttt}
\end{equation}

For this stress-energy tensor and this background metric, we call the
solution to \Deqn{EabTab} $h^\text{S}_{ab}$, for reasons explained
below, and its derivation is given elsewhere \cite{Det01, Det05}.
Here we present the results:
\begin{equation}
  h^\text{S}_{ab} = {}_0h_{ab}^\text{S}+ {}_2h_{ab}^\text{S}
          + {}_3h_{ab}^\text{S},
\label{hS}
\end{equation}
where
\begin{equation}
  {}_0h^\text{S}_{ab}\mbox{d}x^a {\,{\rm d}x}^b = 2\frac{\MU}{r}({\,{\rm d}t}^2 + {\,{\rm d}r}^2)
\label{h0}
\end{equation}
is the Coulomb $\MU/r$ part of the Schwarzschild metric, and
\begin{eqnarray}
  {}_2h^\text{S}_{ab} {\,{\rm d}x}^a {\,{\rm d}x}^b
          &=& \frac{4 \MU}{r} \calE_{ij}x^i x^j {\,{\rm d}t}^2
        - 2 \frac{4\MU r}{3} \epsilon_{kpq} \calB^q{}_i x^p x^i {\,{\rm d}t} {\,{\rm d}x}^k
\nonumber\\ &&
  + \; \text{$\dot\calE_{ij}$ and $\dot\calB_{ij}$ terms}
\label{h2mu}
\end{eqnarray}
 are the quadrupole tidal distortions of the Coulomb part.
The terms involving $\dot\calE_{ij}$ and $\dot\calB_{ij}$ are more
complicated and are not given here. The octupole tidal distortions of
the Coulomb field are
\begin{eqnarray}
 {}_3h^\text{S}_{ab} {\,{\rm d}x}^a {\,{\rm d}x}^b &=&
      \frac{\MU}{3r}\calE_{ijk} x^i x^j x^k
      \left[5 {\,{\rm d}t}^2 + {\,{\rm d}r}^2
             + 2 \sigma_{AB} {{\rm d}} x^A {{\rm d}} x^B \right]
\nonumber \\ & &
    \quad {} - 2 \frac{10\MU}{9r} \epsilon_{kpq}
          \calB^q{}_{ij} x^p x^i x^j {\,{\rm d}t} {\,{\rm d}x}^k.
\label{h3muab}
\end{eqnarray}
Recall that $\sigma_{AB}$ is the two dimensional metric on the
surface of a constant $r$ two sphere.

The perturbation $h^\s_{ab}$ is a solution to \Deqn{EabTab} only in a
neighborhood of $\Gamma$. The next perturbative-order terms which are
not included in $h^\s_{ab}$ scale as $\MU r^3/\calR^4$. The operator
$E_{ab}$ involves second derivatives, and it follows that for
$h^\s_{ab}$ given above
\beq
  E_{ab}(h^s) = -8\pi T_{ab} +\text{O}(\MU r /\calR^4).
\label{Eabhs}
\eeq

In some circumstances we might wish to introduce a window function
$W$ similar to that described in Sect.~\ref{alternative}, which would
multiply all of the terms on the right hand side of Eq.~(\ref{hS}).
If so, the window function near by $\MU$ must be restricted by the
condition that
\begin{equation}
  W = 1 + {\text{O}}(r^4/\calR^4)
\end{equation}
in order to preserve the delicate features of $h^\text{S}_{ab}$ in a
neighborhood of $\MU$, especially the property revealed in
\Deqn{Eabhs}. Away from $\MU$, it is only necessary that $W$ vanish
in some smooth manner.

The perturbations ${}_2h^\text{S}_{ab}$ and ${}_3h^\text{S}_{ab}$
should not be confused with a consequence of Newtonian tides. When a
small Newtonian object moves through spacetime, its mass distribution
is tidally distorted by the external gravitational field. The extent
of this distortion depends upon the size $d$ of the object itself.
For a self-gravitating, non-rotating incompressible fluid\footnote{A
terse but adequate description of perturbative tidal effects on a
Newtonian, self-gravitating, non-rotating, incompressible fluid is
given on p.~467 of \cite{ChandraHHMS}.}, the quadrupole distortion of
the matter leads to a change in the Newtonian gravitational potential
outside the object which scales as $\udelta U \sim {\cal I}_{ij}x^i
x^j/r^3\sim d^{5}/r^{3}\calR^2$, where ${\cal I}_{ij}$ is the mass
quadrupole moment tensor. Such behavior is not at all similar to that
of ${}_2h^\text{S}_{tt} = {\text{O}}(\MU r/\calR^2)$, and
${}_3h^\text{S}_{tt} = {\text{O}}(\MU r^2/\calR^3)$.

 The quadrupole distortion revealed in ${}_2h^\text{S}_{tt}$ is not a
consequence of a distortion of the object $\MU$ itself, but rather
results from the curvature of spacetime acting on the monopole field
of $\MU$ and has no Newtonian counterpart.

\subsection{A small object moving through spacetime}
\label{pert3}

\newcommand{\iN}{{\text{in}}}
\newcommand{\rin}{{r_{\text{in}}}}

As a concrete example we now focus on a small Newtonian object of
mass $\MU$ and characteristic size $d$ moving through some given
external vacuum spacetime with metric $g_{ab}$. Naturally, $\MU$ is
approximately moving along a geodesic $\Gamma$, and $g_{ab}$ has a
characteristic length and time scale $\calR$ associated with
$\Gamma$. We assume that $\MU$ and $d$ are both much smaller than
$\calR$.

In a region comparable to $d$, the object appears Newtonian, and its
gravitational potential can be determined. The structure of the
object depends upon details like the density, type of matter, amount
of rotation and whether it is stationary or oscillating.

The Newtonian object might have a mass quadrupole moment ${\cal
I}_{ij} =\text{O}(\MU d^2)$ perhaps sustained by internal stresses in
the matter itself. Independent of the cause of the quadrupole moment,
the external Newtonian gravitational potential would have a
quadrupole part ${\cal I}_{ij}x^i x^j/r^5 $.

The coupling between a mass quadrupole moment of the small object and
an external octupole gravitational field $\calE_{ijk}x^ix^jx^k$
results in the small acceleration of the center of mass, away from
free-fall, given by \cite{ThorneHartle85, Zhang85}
\beq
   a^i = -\frac{1}{2\MU}\calE^{ijk}{\cal I}_{jk}
\eeq
in either the context of Newtonian physics or of General Relativity.
This tidal acceleration scales as
\beq
   a =\text{O}(d^2/\calR^3).
\label{IfromStress}
\eeq

If our small Newtonian object is actually a nonrotating fluid body
then it would naturally be spherically symmetric except for
distortion caused by an external tidal field such as $\calE_{ij}x^i
x^j$. In that case ${\cal I}_{ij} =\text{O}(d^5/\calR^2)$ as
discussed at the end of Sect.~\ref{pert2} \cite{ChandraHHMS}, and the
tidal acceleration then scales as
\beq
   a =\text{O}(d^5/\MU\calR^5).
\label{IfromTide}
\eeq
We conclude that a Newtonian object in free motion is only allowed an
acceleration away from free-fall which is limited as in
\Deqns{IfromStress} or (\ref{IfromTide}). Any larger acceleration
must involve some non-gravitational force.

It is also possible to analyze the situation if we replace the
Newtonian object with a small Schwarzschild black hole of mass $\MU$.
In that case it is easiest to turn the perturbation problem
inside-out and to consider the Schwarzschild metric as the background
with the metric perturbation being caused by $H_{ab}$ given in
\Deqns{H2} and (\ref{H3}). One boundary condition is that $h_{ab}$
approach $H_{ab}$ for $\MU \ll r \ll \calR$. The boundary condition
at the event horizon is that $h_{ab}$ be an ingoing wave, or
well-behaved in the time independent limit. The time independent problem 
is well studied; 
historically in Refs.~\cite{ReggeWheeler,
Zerilli}, more recently in the present context in Ref.~\cite{Det01}, and with
slow time dependence in Refs.~\cite{Poisson04a, Poisson04b}.

In the time independent limit, the generic quadrupole perturbation of
the metric of the Schwarzschild spacetime results in
\bea
  (g^\schw_{ab} + h^\schw_{ab})\;\mbox{d}x^a \mbox{d}&&\!\!\!\!\!\!x^b
    = - \Big(1-\frac{2\MU}{r}\Big)
         \Big[1-\calE_{ij}x^ix^j\Big(1-\frac{2\MU}{r}\Big)\Big]
          \mbox{d}t^2
\nonumber \\ & &
  {} + \frac{4}{3} \epsilon_{kpq}{\cal B}^q{}_i x^p x^i
                          \Big(1-\frac{2\MU}{r}\Big) \mbox{d}t\, \mbox{d}x^k
  {} + \Big(\frac{1}{1-2\MU/r} - \calE_{ij}x^ix^j\Big) \mbox{d}r^2
\nonumber\\ & &
  {} + \Big[r^2 - \big(r^2-\MU^2\big)\calE_{ij}x^i x^j\Big]
  \big(\mbox{d}\theta^2+\sin^2\theta \mbox{d}\phi^2\big).
\label{pertSchw}
\eea
In this expression $x^i$ represents $x$,$y$ and $z$ which are related
to $r$, $\theta$ and $\phi$ in the usual way in Cartesian space.

It is elementary to check that if $\MU=0$ then this reduces to the
time independent limit of \Deqn{H2}. If $\calE_{ij}$ and
$\calB_{ij}=0$ then this reduces to the Schwarzschild metric. And the
terms which are bilinear in $\MU$ and either $\calE_{ij}$ or
$\calB_{ij}$ are equivalent to the time independent limit of
\Deqn{h2mu}. An expression with similar features holds for the
octupole perturbations.

The metric of \Deqn{pertSchw} represents a Schwarzschild black hole
at rest on the geodesic $\Gamma$ in a time-independent external
spacetime.
 And note that there is no black hole quadrupole moment induced by
the external quadrupole field as there are no quadrupole $1/r^3$
terms in this metric in the region where $\MU \ll r \ll \calR$ .
 The Schwarzschild black hole equivalent of ${\cal I}_{ij}$ vanishes.
It follows that, in this situation, the black hole has no
acceleration away from $\Gamma$.

Time dependence in $\calE_{ij}$ slightly changes this situation. In
\cite{Det01}, it is argued that with slow time dependence, with a
time-scale $\text{O}(\calR)$, the induced quadrupole field of the
Schwarzschild metric in fact scales as $\sim \MU^5/r^3\calR^2$, and
that the acceleration from coupling with an external octupole field,
$\calE_{ijk}x^ix^jx^k \sim r^3/\calR^3$, gives an acceleration
\beq
    a =\text{O}(\MU^4/\calR^5).
\eeq
This result is consistent with the Newtonian result in
\Deqn{IfromTide} if the size $d$ of the Newtonian object is replaced
with the mass $\MU$ of the black hole.

An elementary approach using dimensional analysis arrives at this same
result. Acceleration is a three-vector with a unit of 1/length. If the
only quantities in play are $\MU$, $\calE_{ij}$ and $\calE_{ijk}$. The
only combination of these which yields a vector with the units of
acceleration is $\MU^4\calE^{ijk} \calE_{jk} =\text{O}(\MU^4/\calR^5)$.

However, Eric Poisson has pointed out that a combination involving the
magnetic multipole moments, such as $\MU^3\calE^{jk}
\calB_{kl}\epsilon^l{}_{ji} =\text{O}(\MU^3/\calR^4)$, might provide a
lower order acceleration.


The field $h^\s_{ab}$ is now seen to satisfy the requirements desired
for a ``Singular field:''
 \begin{enumerate}
   \item[A.] $h^\s_{ab}$ is a solution of the field equation in
       the vicinity of a $\delta$-function mass source on a
       geodesic $\Gamma$.
   \item[B.] $h^\s_{ab}$ exerts no force back on its $\delta$-function
       source as evidenced by the facts that $h^\s_{ab}$ is the part
       of the perturbed Schwarzschild geometry that is linear in
       $\MU$, and that the small black hole has acceleration no larger
       than $\text{O}(\MU^3/\calR^4)$, while all that is required is
       that the acceleration be no larger than
       $\text{O}(\MU^2/\calR^3)$.
 \end{enumerate}

\section{Self-force from gravitational perturbation theory}
\label{gsf1stOrder}

For an overview of the general approach to gravitational self-force
problems about to be described, we refer back to the treatment of the
electromagnetic self-force in Sect.~\ref{E-M}, the toy-problem of
Sect.~\ref{toyproblem}, and particularly to the introduction of
$h^\text{S}_{ab}$ in Sects.~\ref{pert2} and \ref{pert3} .

At a formal level, we begin with a metric $g_{ab}$ which is a vacuum
solution of the Einstein equation and look for an approximate solution
for $h^\act_{ab}$ from
\begin{equation}
  G(g + h^\act) = 8\pi T + {\text{O}}(h^2),
\label{1stAct}
\end{equation}
with appropriate boundary conditions, where $T_{ab} =
{\text{O}}(\MU)$ is the stress-energy tensor of a point particle
$\MU$.

Initially we assume that $\MU$ is moving along a geodesic $\Gamma$.
In a neighborhood of $\Gamma$, $h_{ab}$ is well approximated by
$h^\text{S}_{ab}$. Thus we define $h^\text{R}_{ab}$ via the
replacement
\begin{equation}
  h^\act_{ab} = h^\text{S}_{ab} + h^\text{R}_{ab},
\label{hact}
\end{equation}
and use the expansion in \Deqn{G0+G1+G2} and the definition in
\Deqn{Eab} to write
\begin{eqnarray}
  G_{ab}(g+h^\act) &=& G_{ab}(g) - E_{ab}(h^\act) + {\text{O}}(h^2)
\nonumber\\
       &=& - E_{ab}(h^\text{R}) - E_{ab}(h^\text{S}) + {\text{O}}(h^2)
\end{eqnarray}
where we use the assumption that $G_{ab}(g)=0$ and the linearity of
the operator $E_{ab}(h)$.

In Sect.~\ref{pert2} the properties of $h^\text{S}_{ab}$ were chosen
carefully so that
\begin{equation}
   E_{ab}(h^\text{S}) = - 8\pi T_{ab}
                + {\text{O}}(\MU r/\calR^4)\text{ in a neighborhood of
   $\MU$.}
\label{GhS}
\end{equation}
We can demonstrate this result by letting ${}_4h^\text{S}_{ab} =
{\text{O}}( \MU r^3 / \calR^4)$ be the next term not included in the
expansion (\ref{hS}). The operator $E_{ab}$ has second order spatial
derivatives, and every time derivative brings in an extra factor of
$1/\calR$. Thus $E_{ab}({}_4h^\text{S}_{ab}) = {\text{O}}(\MU
r/\calR^4)$, and \Deqn{GhS} follows.

Now we define the \textit{effective source}
\begin{eqnarray}
   8\pi S_{ab} &\equiv& 8\pi T_{ab} + E_{ab}(h^\text{S}),
\nonumber\\
                &=& {\text{O}}(\MU r/\calR^4).
\label{Sdef}
\end{eqnarray}
Thus $S_{ab}$ is zero at $r=0$, where it is continuous but not
necessarily differentiable. Everywhere else $S_{ab}$ is $C^\infty$.

The first perturbative order problem \Deqn{1stAct} is now reduced to
solving
\begin{equation}
   E_{ab}(h^\text{R}) = - 8\pi S_{ab},
\label{1stOrdEq}
\end{equation}
and then \Deqn{hact} reconstructs $h^\act_{ab}$. The limited
differentiability of $S_{ab}$ causes no fundamental difficulty for
determining $h^\text{R}_{ab}$, and introduces no small length scale
either. The resulting $h^\text{R}_{ab}$ will be $C^2$ at the location
of the point mass, and $C^\infty$ elsewhere.

At this order of approximation Sect.~\ref{pert3} showed that the mass
$\MU$ moves along a geodesic of the \textit{actual} metric
$g^\act_{ab}$ with $h^\s_{ab}$ removed, {\it i.e.} along a geodesic
of $g_{ab} + h^\act_{ab}- h^\s_{ab} = g_{ab} + h^\R_{ab}$.
Thus, the gravitational self-force results in geodesic motion not in
$g_{ab}$ but rather in $g_{ab}+h^\R_{ab}$.

Admittedly, $g_{ab} + h^\R_{ab}$ is not truly a vacuum solution of the
Einstein equation. But, by construction it is clear that
\beq
  G_{ab}(g+h^\R) = {\text{O}}(\MU r/\calR^4).
\label{Gg+hs}
\eeq
More terms of higher order in $r/\calR$ in the expression for
$h^\s_{ab}$ would result in a remainder with more powers of $r/\calR$
on the right hand side of \Deqn{Gg+hs}. But these would not change
the first derivatives of $h^\R_{ab}$ on $\Gamma$ which are all that
would appear in the geodesic equation for $\MU$. So the expansion for
$h^\s_{ab}$ as given in Sect.~\ref{pert3} is adequate for our
purposes.

\subsection{Dissipative and conservative parts}
\label{gsfDisCon}

When viewed from near by, the effect of the gravitational self-force on
a small mass $\MU$ arises as a consequence of the purely local
phenomenon of geodesic motion. In the neighborhood of $\MU$, it is
impossible then to distinguish the dissipative part of the self-force
from the conservative part.

Viewed from afar with the usually appropriate boundary conditions, the
metric perturbation $h^\act_{ab}$ is actually the retarded field
$h^\ret_{ab}$ and it is often useful then to distinguish the
dissipative effects which remove energy and angular momentum from the
conservative effects which might affect, say, the orbital frequency.

In the case that $h^\act_{ab} = h^\ret_{ab}$, it is natural to define
the dissipative part of the regular field as
\beq
  h^\dis_{ab} = \frac{1}{2}(h^\ret_{ab}- h^\adv_{ab})
\eeq
The advanced and the retarded fields are each solutions of the same wave
equation with the same $\delta$-function source. Thus their
 difference is a solution of the homogeneous wave equation and is
therefore regular at the point mass. And the dissipative effects of the
self-force are revealed as geodesic motion in the metric
$g_{ab}+h^\dis_{ab}$.

In a complementary fashion, the conservative part of the regular
field is naturally defined as
\bea
  h^\con_{ab} &=& h^\R_{ab} - \frac{1}{2}(h^\ret_{ab}- h^\adv_{ab})
\nonumber\\
   &=& h^\ret_{ab} - h^\s_{ab} - \frac{1}{2}(h^\ret_{ab}- h^\adv_{ab})
\nonumber\\
   &=& \frac{1}{2}(h^\ret_{ab} + h^\adv_{ab}) - h^\s_{ab}
\eea
And the conservative effects of the self-force are revealed as
geodesic motion in the metric $g_{ab}+h^\con_{ab}$.

With these definitions it is natural that
\beq
  h^\R_{ab} = h^\con_{ab} + h^\dis_{ab} .
\eeq

This decomposition into conservative and dissipative parts follows an
aspect of the procedure that Mino describes \cite{Mino03} as a possible
method for computing the dissipative effects of gravitational radiation
reaction on the Carter constant
\cite{Carter68a,Carter68b} for a small mass orbiting a Kerr black
hole.

\subsection{Gravitational self-force implementations}
\label{gsfImplementation}

When it is actually time to search for some self-force consequences
there are a number of different choices to be made.

\subsubsection{Field regularization via the effective source}

The majority of this review has been leading toward a natural
implementation of self-force analysis using the standard 3+1
techniques of numerical relativity. Assume that $h^\R_{ab}$ and its
first derivatives, and also the position and four-velocity of $\MU$
are known at one moment of time.
\begin{enumerate}
\item Use the position and four-velocity of $\MU$ to analytically
    determine $h^\s_{ab}$.

\item Obtain the effective source $S_{ab}$ via \Deqn{Sdef}.

\item Evolve \Deqn{1stOrdEq} for $h^\R_{ab}$ one step forward in
    time.

\item Move the particle a step forward in time using the geodesic
    equation for $g_{ab}+h^\R_{ab}$.
\item Repeat.
\end{enumerate}
Section \ref{fieldreg} describes the application of this approach to
a scalar field problem and includes figures which reveal some generic
characteristics of the source function.

\subsubsection{Mode-sum regularization}

Mode-sum regularization \cite{Barack01, BarackOri02} avoids the
singularity of $h^\act_{ab}$ and its derivatives on $\Gamma$ by an
initial multipole-moment decomposition, say, into spherical harmonic
components $h^{\act\ell m}_{ab}$.
 With the assumption that $h^\text{S}_{ab}$ is carefully defined away from
$\MU$ in a fashion that also allows for a decomposition in terms of
spherical harmonics $h^{\text{S}\ell m}_{ab}$, then $h^{\text{R}\ell
m}_{ab} = h^{\act\ell m}_{ab} - h^{\text{S}\ell m}_{ab}$ would be the
decomposition of $h^\text{R}_{ab}$. The collection of the multipole
moments $h^{\text{S}\ell m}_{ab}$, their derivatives and various of
their linear combinations are, together, known as ``regularization
parameters.'' This essentially leads to the mode-sum regularization
procedure of Barack and Ori\cite{Barack01, BarackOri02} which has been
used in nearly all of the self-force calculations to date.

\subsubsection{The gravitational self-force actually resulting in
acceleration}

We have strongly pushed our agenda of treating the gravitational
self-force in local terms as geodesic motion through a vacuum spacetime
$g_{ab}+h^\R_{ab}$. However, when viewed from afar the worldline
$\Gamma$ of $\MU$ is indeed accelerated and not a geodesic of the
background geometry $g_{ab}$. This acceleration can be described as a
consequence of $\MU$ interacting with a spin-2 field $h^\R_{ab}$ which
leads to the resulting acceleration
\begin{equation}
   u^b \nabla_b u^a = - \left(g^{ab}+ u^a u^b\right) u^c u^d
     \left(\nabla_{c} h^\text{R}_{db} - \frac12 \nabla_b h^\text{R}_{cd}\right)
 \label{gravforce-a}
\end{equation}
away from the original worldline in the original metric $g_{ab}$.

Under some circumstances this might be a convenient interpretation.
The resulting worldline would be identical to the geodesic of
$g_{ab}+h^\text{R}_{ab}$ and would correctly incorporate all
self-force effects, although the worldline would not be parameterized
by the actual proper time. It is important to note that the
acceleration of \Deqn{gravforce-a} cannot be measured with an
accelerometer and, by itself, has no actual, direct physical
consequence.

In the next section we describe some general consequences of gauge
transformations in perturbation theory. Be warned that if
\Deqn{gravforce-a} is used to calculate the deviation $\zeta^a$ of
the worldline away from a geodesic in the background metric $g_{ab}$,
then any gauge transformation whose gauge vector $\xi^a = -\zeta^a$,
on the world line, would automatically set the right hand side of
\Deqn{gravforce-a} to zero and leave $\MU$ on its original geodesic.
This possibility certainly confuses the interpretation of the right
hand side of \Deqn{gravforce-a}. Such a removal of the self-force
only works as long as the deviation vector $\zeta^a \sim
{\text{O}}(h)$. If self-force effects accumulate in time, such as
from dissipation or orbital precession, then after a long enough time
the effects of the self-force will be revealed.

\section{Perturbative gauge transformations}
\label{gauge}

\newcommand{\prim}[1]{{#1^\prime}}

In General Relativity, the phrase ``choice of gauge'' has different
possible interpretations depending upon whether one is interested in
perturbation theory or, say, numerical relativity. With numerical
relativity, ``choice of gauge'' usually refers to the choice of a
specific coordinate system, with the understanding that general
covariance implies that the meaning of a calculated quantity might be
as ambiguous as the coordinate system in use.

In perturbation theory the ``choice of gauge'' is more subtle. One
considers the difference between the actual metric ${g}^\act_{ab}$ of
a spacetime of interest and an abstract metric ${{g}}_{ab}$ of a
given, background spacetime. The difference
\begin{equation}
  h_{ab} = {g}^\act_{ab} - g_{ab}
 \label{hdef}
\end{equation}
is assumed to be small. The perturbed Einstein equations govern
$h_{ab}$, and knowing $h_{ab}$ might provide answers to questions
concerning the propagation and emission of gravitational waves, for
example.

In this perturbative context ``choice of gauge'' involves the choice
of coordinates, but in a very precise sense \cite{Sachs64,
StewartWalker74, Bardeen80, BarackOri01}. The subtraction in
\Deqn{hdef} is ambiguous. The two metrics reside on different
manifolds, and there is no unique map from the events on one manifold
to those of another. Usually the names of the coordinates are the
same on the two manifolds, and this provides an implicit mapping
between the manifolds. But this mapping is not unique. For example,
the Schwarzschild geometry is spherically symmetric. This allows the
Schwarzschild coordinate $r$ to be defined in terms of the area $4\pi
r^2$ of a spherically symmetric two-surface. The perturbed
Schwarzschild geometry is not spherically symmetric, and to describe
the coordinate $r$ on the perturbed manifold as the ``Schwarzschild
$r$'' does not describe the meaning of $r$ in any useful manner and
is not a perturbative choice of gauge.

In perturbation theory a gauge transformation is an infinitesimal
coordinate transformation of the perturbed spacetime
\begin{equation}
  x_\new^a = x^a_\old + \xi^a , \quad \text{where} \quad \xi^a = {\text{O}}(h),
 \label{gaugetransformation}
\end{equation}
and the coordinates $x^a_\new$, $x^a_\old$, and the coordinates on
the abstract manifold are all described by the same names, for
example $(t,r,\theta,\phi)$ for perturbations of the Schwarzschild
geometry.
 The transformation of \Deqn{gaugetransformation} not only changes
the components of a tensor by ${\text{O}}(h)$, in the usual way, but
also changes the mapping between the two manifolds and hence changes
the subtraction in \Deqn{hdef}. With the transformation
(\ref{gaugetransformation}),
\begin{eqnarray} 
  h^\text{\scriptsize new}_{ab}
    &=&
     \left({{g}}_{cd} + h^\text{\scriptsize old}_{cd}\right)
         \frac{\partial x_\old^c}{\partial x_\new^a}
         \frac{\partial x_\old^d}{\partial x_\new^b}
   - \left( {{g}}_{ab}
       + \xi^c \frac{\partial {{g}}_{ab}}{\partial x^c}\right) .
\end{eqnarray}
The $\xi^c$ in the last term accounts for the ${\text{O}}(h)$ change
in the event of the background used in the subtraction. After an
expansion, this provides a new description of $h_{ab}$
\begin{eqnarray}
   h^\text{\scriptsize new}_{ab} &=& h^\text{\scriptsize old}_{ab}
     - {{g}}_{cb} \frac{\partial \xi^c}{\partial x^a}
     - {{g}}_{cb} \frac{\partial \xi^d}{\partial x^b}
     - \xi^c \frac{\partial {{g}}_{ab}}{\partial x^c}
\nonumber \\ &=& h^\text{\scriptsize old}_{ab}
     -\pounds_{\xi} {{g}}_{ab}
  = h^\text{\scriptsize old}_{ab} - 2 \nabla_{(a} \xi_{b)}
\label{hprime}
\end{eqnarray}
through ${\text{O}}(h)$; the symbol $\pounds$ represents the Lie
derivative and $\nabla_{a}$ is the covariant derivative compatible
with $g_{ab}$. A gauge transformation does not change the actual
perturbed manifold, but it does change the coordinate description of
the perturbed manifold.

A little clarity is revealed by noting that
\begin{equation}
 E_{ab}(\nabla_{(c} \xi_{d)}) \equiv 0
\label{EabGauge}
\end{equation}
for any $C^2$ vector field $\xi^a$; and if $\xi^a$ has limited
differentiability or is a distribution, then \Deqn{EabGauge} holds in
a distributional sense \cite{Det05}. Thus $-2\nabla_{(a} \xi_{b)}$ is
a homogeneous solution of the linear \Deqn{Eab}. It appears as though
any $-2\nabla_{(a} \xi_{b)}$ may be added to an inhomogeneous
solution of \Deqn{Eab} to create a ``new'' inhomogeneous solution. In
fact the new solution is physically indistinguishable from the
old---they differ only by a gauge transformation with gauge vector
$\xi^a$.

Generally, the four degrees of gauge freedom contained in the gauge
vector $\xi^a$ are used to impose four convenient conditions on
$h_{ab}$. For perturbations of the Schwarzschild metric, it is common
to use the Regge Wheeler gauge which sets four independent parts of
$h_{ab}$ to zero; this results in some very convenient algebraic
simplifications. The Lorenz gauge requires that
$\nabla_a(h^{ab}-\frac12g^{ab}h^c{}_c)=0$ and is formally attractive
but unwieldy in practice \cite{barack:01, DetPoisson04,
BarackSago07}.

The Bianchi identity implies that there are four relations among the
ten components of the Einstein equations. Choosing a gauge helps
focus on a self-consistent method for solving a subset of these
equations.  
A physicist might have a favorite for a gauge choice, but Nature has
no preference whatsoever.

\section{Gauge confusion and the gravitational self-force}
\label{gaugeconfusion}

\newcommand{\ubar}{{\bar u}}
\newcommand{\rdot}{{\dot R}}
\newcommand{\rddot}{{\ddot R}}
\newcommand{\Lie}{{\cal L}}

If a particular physical consequence of the gravitational self-force
requires a particular choice of gauge, then it is unlikely that this
physical consequence has any useful interpretation. This was already
demonstrated with the example presented in Sect.~\ref{newtonian}
where the magnitude of the effect of the Newtonian self-force on the
period in an extreme-mass-ratio binary depended upon the definition
of the variable $r$.

The quasi-circular orbits of the Schwarzschild geometry provide a fine
example which reveals the insidious nature of gauge confusion in
self-force analyses. Ref. \cite{Det08} contains a thorough discussion
of this subject and this section has two self-force examples which
highlight the confusion that perturbative gauge freedom creates.

It is straightforward to determine the components of the geodesic
equation for the metric $g^\schw_{ab} + h^\R_{ab}$. A consequence of
these is that the orbital frequency of $\MU$ in a quasi-circular orbit
about a Schwarzschild black hole of mass $M$ is given by
\begin{equation}
  \Omega^2
     = \frac{{M} }{{r}^3}
        - \frac{{r}-3{M} }{2 {r}^2} u ^a u ^b \partial_r h^\R_{ab}
\label{Omega2}
\end{equation}
which can be proven to independent of the gauge choice. Clearly the
self-force makes itself known to the orbital frequency through the
last term. So we focus on the orbit at radius $r=10M$, choose to
work in the Lorenz gauge, work hard and successfully evaluate all of
the components of the regularized field $h^\R_{ab}$ as well as its
radial derivative. Then we calculate the second term in \Deqn{Omega2}
and determine that $\Omega$ changes by a specific amount $\Delta
\Omega_\text{lz}$. We now know the gauge invariant change in the
orbital frequency for $\MU$ in the orbit at $10M$.

Or do we? To check this result we repeat the numerical work but this
time use the Regge-Wheeler gauge, and find that the change in
$\Omega$ is $\Delta \Omega_\text{rw}$ and
\beq
  \Delta \Omega_\text{rw} \ne \Delta \Omega_\text{lz}\;!
\eeq

What's going on? For a quasi circular orbit $\Omega$ can be
\textit{proven} to be independent of gauge, and yet with two
different gauges we find two different orbital frequencies for the
single orbit at $10M$.

When I first discovered this conundrum I was reminded of my experience
trying to understand special relativity and believing that apparently
paradoxical situations made special relativity logically inconsistent.
Eventually the paradoxes vanished when I understood that coordinates
named $t$, $x$, $y$ and $z$ are steeped in ambiguity and that only
physical observables are worth calculating and discussing.

The resolution of this self-force confusion is similar. The two
evaluations of $\Omega^2$ are each correct. But, one is for the orbit
at the Schwarzschild radial coordinate $r=10M$ in the Lorenz gauge,
while the other is at the Schwarzschild radial coordinate $r=10M$ in
the Regge-Wheeler gauge. These are two distinct orbits. In fact, the
gauge vector $\xi^a$ which transforms from the Lorentz gauge to the
Regge-Wheeler gauge has a radial component $\xi^r$ whose magnitude is
just right to make the change in the first term in \Deqn{Omega2}
balance the change in the second term.

The angular frequency of $\MU$ orbiting a black hole is a physical
observable and independent of any gauge choice. But the perturbed
Schwarzschild geometry is not spherically symmetric and there is then
no natural definition for a radial coordinate.

A second example of gauge confusion appears when one attempts to find
the self-force effect on the rate of inspiral of a quasi-circular
orbit of Schwarzschild. It is natural to find the energy $E$,
$\Omega$ and $dE/dt$ all as functions of the radius of a circular
orbit and then to use
\begin{equation}
  \frac{d\Omega}{dt} = \frac{dE}{dt}
           \times\frac{d\Omega/dr}{dE/dr}
\label{dOmegadt}
\end{equation}
to determine the rate of change of $\Omega$. We can find the self
force effect on each of these quantities so we can apparently find
the self force effect on $d\Omega/dt$ which is a physical observable
and must be gauge invariant.

This situation is subtle. Why do we believe \Deqn{dOmegadt}? With some
effort it can be shown that the geodesic equation for
$g^\schw_{ab}+h^\R_{ab}$ implies that \Deqn{dOmegadt} holds for a
quasi-circular orbit \cite{Det08}. Part of this proof depends upon the
$t$-component of the geodesic equation which is
\begin{eqnarray}
  \frac{dE}{dt} &=& -\frac{1}{2 u^t} u^a u^b\partial_t h^\R_{ab} ,
\label{dEdt}
\end{eqnarray}
and this is a gravitational self-force effect. But, note that the
right hand side of \Deqn{dOmegadt} is already first order in
$h^\R_{ab}$ from the factor $dE/dt$. While self-force effects on
$d\Omega/dr$ and $dE/dr$ can be found, if these are included then
second order self force effects on $dE/dt$ must also be found for a
consistent solution.

The end result is that you really can't see the effect of the
conservative part of the self-force on the waveform for quasi-circular
orbits using first order perturbation theory.

\section{Steps in the analysis of the gravitational self-force}
\label{steps}

We now highlight the major steps involved in most gravitational
self-force calculations.

First the metric perturbation $h^\act_{ab}$ is determined. For a
problem in the geometry of the Schwarzschild metric, this involves
solving the
Regge-Wheeler\cite{ReggeWheeler} and the Zerilli\cite{Zerilli}
equations to determine the actual metric perturbations. The Kerr
metric still presents some challenges. The Teukolsky
\cite{Teukolsky73, SasakiNakamura} formalism can provide the Weyl
scalars but finding the metric perturbations \cite{Whiting:2005hr} from
these is difficult at best, and does not include the non-radiating
monopole and dipole perturbations. One possibility for Kerr is to find
the metric perturbations directly, perhaps in the Lorenz gauge, but
this would likely require a $3+1$ approach. Another possibility being
discussed \cite{Barack:2007we} is to Fourier transform in $\phi$, and
then use a $2+1$ formalism which results in an $m$-sum. Rotating black
holes continue to be a challenge for self-force calculations.

Next, the singular field $h^\text{S}_{ab}$ is identified for the
appropriate geodesic in the background spacetime.
A general expansion of the singular field is available \cite{Det01}, but
it is not elementary to use.\footnote{\label{foot} Expansions for the somewhat related
``direct'' field are also available \cite{BarackOri02, Mino97,
QuinnWald97, Mino02, BMNOS02, BarackOri03, Poisson05}, though their use
is, similarly, not at all elementary.}
Work in progress 
provides a constructive procedure for the THZ coordinates in the
neighborhood of a geodesic, and this would lead to explicit
expressions for $h^\text{S}_{ab}$ in the natural coordinates of the
manifold. However, this procedure is not yet in print, and it is not
yet clear how difficult it might be to implement.

Then the perturbation is regularized by subtracting the singular field
from the actual field resulting in $h^\text{R}_{ab} = h^\act_{ab} -
h^\text{S}_{ab}$. Most applications have taken this step using the
mode-sum regularization procedure of Barack and Ori\cite{Barack01,
BarackOri02}. In this case, a mode-sum decomposition of the singular (or
``direct,'' cf. footnote \ref{foot}) field is identified and then removed from the mode-sum
decomposition of the actual field. The remainder is essentially the
mode-sum decomposition of the regular field. Generally, this mode-sum
converges slowly as a power law in the mode index, $l$ or $m$. Although
some techniques have been used to speed up this convergence
\cite{DetMessWhiting03}. More recently, ``field regularization''
(discussed in Sect.~\ref{fieldreg} and in \cite{VegaDet08}) has been
used for scalar field self-force calculations. For this procedure in
the gravitational case,
\Deqn{1stOrdEq} might be used to obtain the regular field $h^\text{R}_{ab}$
directly via $3+1$ analysis.

After the determination of $h^\text{R}_{ab}$, the effect of the
gravitational self-force is then generically described as resulting
in geodesic motion for $\MU$ in the metric $g^{\mbox{\scriptsize
o}}_{ab}+h^\text{R}_{ab}$. This appears particularly straightforward
to implement using field regularization. Alternatively, the motion
might also be described as being accelerated by the gravitational
self-force as described in \Deqn{gravforce-a}.

At this point, one should be able to answer the original
question---whatever that might have been! In fact, the original
question should be given careful consideration \textit{before}
proceeding with the above steps. Formulating the question might be as
difficult as answering it. It is useful to keep in mind that only
physical observables and geometrical invariants can be defined in a
manner independent of a choice of coordinates or a choice of
perturbative gauge.

My prejudices about the above choices for each step are not well
hidden. But, for whatever technique or framework is in use, a
self-force calculation should have the focus trained upon a physical
observable, not upon the method of analysis.

Self-force calculations unavoidably involve some subtlety. Experience
leads me to be wary about putting trust in my own unconfirmed results.
Good form requires independent means to check analyses. Comparisons
with the previous work of others, with Newtonian and post-Newtonian
analyses, or with other related analytic weak-field situations all lend
credence to a result.

\section{Applications}
\label{examples}

Recently, the effect of the gravitational self-force on the orbital
frequency of the innermost stable circular orbit of the Schwarzschild
geometry has been reported by Barack and Sago \cite{BarackSago09}.
They find that the self-force changes the orbital frequency of the
ISCO by $0.4870(\pm0.0006)\MU/M$. To date this result is by far the
most interesting gravitational self-force problem that has been
solved. But it is too recent a result to be described more fully
herein.

\subsection{Gravitational self-force effects on circular orbits of
the Schwarzschild geometry}
\label{deluT}

As an elementary example we consider a small mass $\MU$ in a circular
orbit about the Schwarzschild geometry. Details of this analysis may
be found in \cite{Det08}. The gravitational self-force affects both
the orbital frequency $\Omega$ and also the Schwarzschild
$t$-component of the four-velocity, $u^t$, which is related to a
redshift measurement. The self-force effects on these quantities are
known to be independent of the gauge choice for $h_{ab}$, as would be
expected because they can each be determined by a physical
measurement. However the radius of the orbit depends upon the gauge
in use and has no meaning in terms of a physical measurement.

\begin{figure}
 \centering
 \includegraphics[scale=0.4]{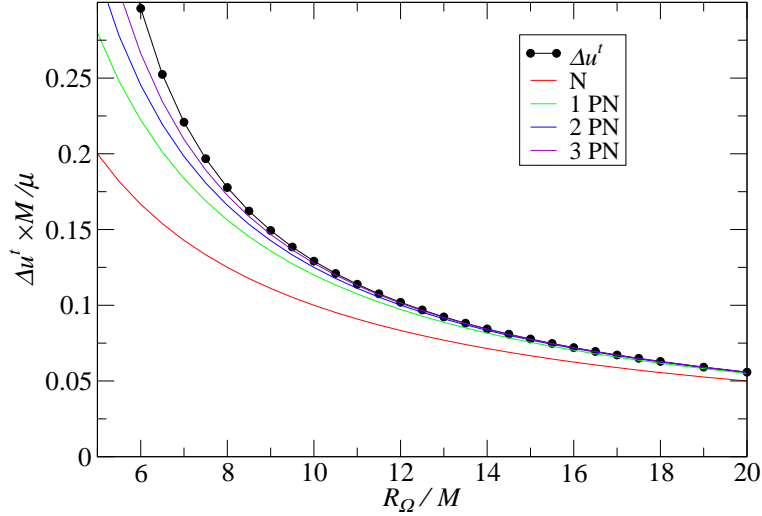}
  \caption{From \cite{Det08}. The quantity ${}_1u^t$, which is the
    gauge independent ${\text{O}}(\MU)$ part of $u^t$, is given as a function of
    $R_{\Omega}$ for circular orbits in the Schwarzschild geometry.
    Also shown are $u_1^t$ as calculated with Newtonian, 1PN and 2PN
    analyses in \cite{Det08} based upon results in
    \cite{BlanchetFayePonsot98} and \cite{BlanchetLR06}. the 3PN line
    is based on a numerical determination of the 3PN
    coefficient in \Deqn{PNexpan} in \cite{Det08}.
  }
 \label{fig:deluT}
\end{figure}

Notwithstanding the above, we define $R_\Omega$ via
\begin{equation}
\Omega^2 = M/R_\Omega^3
\end{equation}
as a natural radial measure of the orbit which inherits the property of
gauge independence from $\Omega$. The quantity $u^t$ can be divided
into two parts $u^t = {}_0u^t + {}_1u^t$, where each part is separately
gauge independent.
 Further the functional relationships between $\Omega$, ${}_0u^t$ and
$R_\Omega$ are identical to their relationships in the geodesic
limit,
\begin{equation}
  {}_0u^t = [1-3(\Omega M)^{2/3}] + O(\MU^2)
\end{equation}
and shows no effect from the self-force. The remainder
\begin{equation}
  {}_1u^t = u^t - {}_0u^t
\end{equation}
is, a true consequence of the self-force, and we plot the numerically
determined ${}_1u^t$ as a function of $R_\Omega$ in
Fig.~\ref{fig:deluT}. The numerical data of Fig.~\ref{fig:deluT} have
also been carefully compared with and seen to be in agreement with the
numerical results of Sago and Barack, as shown in
\cite{SagoBarackDet09}, despite the fact that very different gauges
were in use and different numerical methods were employed.

We have derived a post-Newtonian expansion for ${}_1u^t$ based upon the
work of others \cite{BlanchetFayePonsot98, BlanchetLR06}. Our
expansion is in powers of $m/R_\Omega$, which is $v^2/c^2$ in the
Newtonian limit, and we find
\begin{equation}
  {}_1u^t = \frac{\MU}{M} \left[
    - \left(\frac{M}{R_\Omega}\right)
    - 2 \left(\frac{M}{R_\Omega}\right)^2
    - 5 \left(\frac{M}{R_\Omega}\right)^3 + \cdots \right] ,
\label{PNexpan}
\end{equation}
which includes terms of order $v^6/c^6$. Further, with numerical
analysis we have fit these results to determine a 3PN parameter of
order $v^8/c^8$ and found that the coefficient of the
$({M}/{R_\Omega})^4$ term is $-27.61\pm .03$.

Work in progress, with Blanchet, Le Tiec, and Whiting, includes a full
3PN determination of the same 3PN coefficient as well as a more precise
numerical determination via self-force analyses. The consistency of
these two efforts has the possibility of giving greatly increased
confidence in the self-force numerical analysis as well as in the
post-Newtonian analysis, each of which involves substantial
complications.

This self-force result is primarily only of academic interest. But it
is consistent with a post-Newtonian expansion and includes an estimate
of the previously unknown ${\text{O}}(v^8/c^8)$ coefficient in the
expansion. Modest though it might be, this is a result.

\subsection{Field regularization via the effective source} \label{fieldreg}

The ultimate goal of self-force analysis has become the generation of
accurate gravitational waveforms from extreme mass-ratio inspiral
(EMRI). It would be amusing to ``see'' numerically the waves emitted
by a small black hole in a highly eccentric orbit about a much larger
one and to see the changes in the orbit while the small hole loses
energy and angular momentum.

Such a project appears to require a method to solve for the
gravitational waves while simultaneously modifying the worldline of the
small hole as it responds to the gravitational self-force. The toy
problem in Sect.~\ref{toyproblem} shows how this might be done using
the expertise of numerical relativity groups coupled with the
self-force community.

\begin{figure}[b]
  \centering
  \includegraphics[scale = 1.2,angle=0]{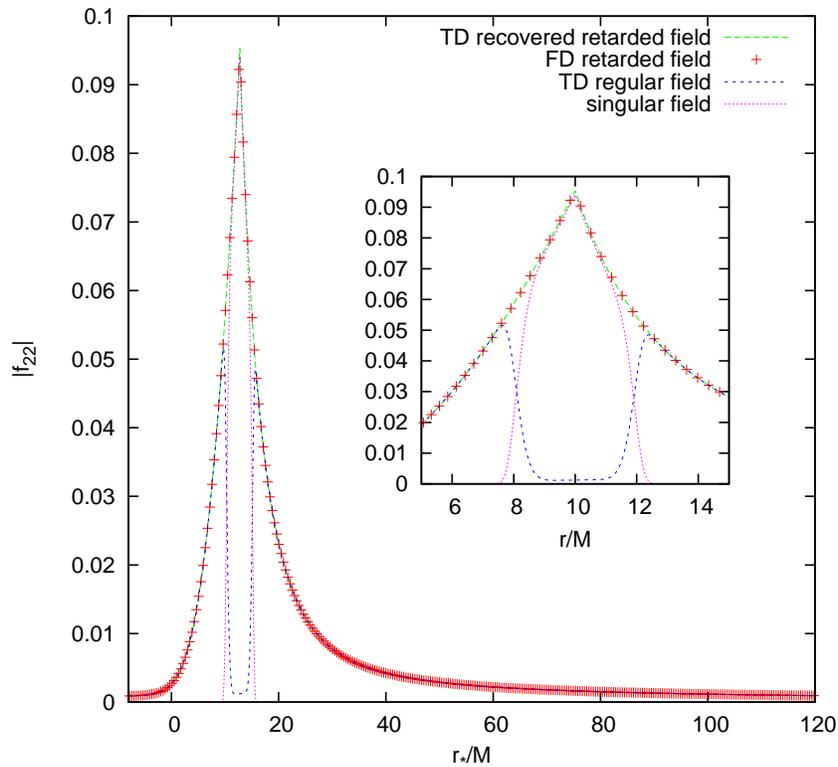}
  \caption{From \cite{VegaDet08}. Comparison of time-domain (TD) and
  frequency-domain (FD) results for the $l=m=2$ multipole moment of the
  scalar field. The regular field is represented by the blue dashed
  line. Adding this to the $l=m=2$ multipole moment of the
  analytically known singular field, $W\psi^S$, results in the
  computed, actual field to good agreement. The inset shows near
  the point charge that $\psi^\text{R}$ is very well behaved and that
  $\psi^\text{R}+W\psi^S$ is indistinguishable from the
  actual, retarded field $\psi^\ret$, just as it should be.
  }
 \label{TDFDcomp}
\end{figure}

Our group is in the early stages of development of infrastructure
that any numerical relativity group could use to get gravitational
self-force projects up and running with a minimum of effort. We
intend to provide the software that will produce the
regularized-field source $S_{ab}$, for a small mass $\MU$ as a
function of location and four-velocity. A numerical relativist could
then evolve the linear field equation
\begin{equation}
   E_{ab}(h^\text{R}) = -8\pi S_{ab} 
\label{GhS1}
\end{equation}
for $h^\text{R}_{ab}$, while simultaneously adjusting the worldline
according to \Deqn{gravforce-a}.

As described in Sect.~\ref{gsf1stOrder} such a computation of
$h^\text{R}_{ab}$ would provide not only the effects of the
gravitational self-force but also the gravitational wave itself.

Ian Vega \cite{VegaDet08} has led a first attempt at directly solving
for the regularized field and self-force using a well tested problem
involving a scalar charge in a circular orbit of the Schwarzschild
geometry. This analysis used a multipole decomposition of the source
and field. And Vega solved for the multipole components in the time
domain using a $1+1$ code.
 Figure \ref{TDFDcomp} shows the $\ell=m=2$ mode and compares the
accurate frequency domain evaluation of the retarded field
$\psi^\ret$ to the sum $\psi^\text{S}+ \psi^\text{R}$ as determined
using $1+1$ methods with field-regularization as described in
Sect.~\ref{toyproblem}.
 Table \ref{tab:only} compares the numerical results of regularized
fields and forces from the field-regularization approach of
\cite{VegaDet08} with the mode-sum regularization procedure
\cite{BarackOri02, BMNOS02} used in \cite{DetMessWhiting03}.

\begin{figure}[b]
  \centering
  \includegraphics[scale=1]{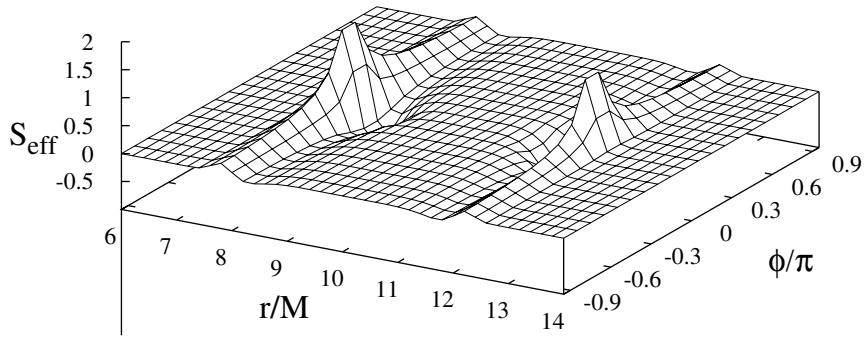} 
  \caption{From \cite{VegaDet08}. The effective source
    $S_{\text{eff}}$ on the equatorial plane for a scalar charge in a
    circular orbit of the Schwarzschild metric. The particle is at
    $r/M=10$, $\phi/\pi=0$, where $S_{\text{eff}}$ appears to have no
    structure on this scale. The spiky appearance is solely a consequence of
    the grid resolution of the figure. In fact the source is
    $C^\infty$ everywhere except at the location of the scalar charge
    where $S_{\text{eff}}$ appears quite calm on this scale.
  }
  \label{figS}
\end{figure}

Figure \ref{figS} shows an example of the source-function used in a
test of this approach with a scalar field. The ``double bump'' shape
far from the charge is a characteristic of any function similar to
$\nabla^2(W/|\vec r-\vec r_0|)$ with a window function $W$ which
satisfies the three window properties given in Sect.~\ref{alternative}.

\def\tstrut{\vrule height3ex depth1.2ex width0pt}%
\begin{table}[t]
\caption{From \cite{VegaDet08}. Summary of scalar field self-force
results for a circular orbits at $R=10M$ and $R=12M$. The error is
determined by a comparison with an accurate frequency-domain
calculation \cite{DetMessWhiting03}.} \label{tab:only} \centering
\begin{tabular}{c c c c c}
\hline\hline
{}{\vrule height3ex depth1.2ex width0pt} & $R$ & Time-domain & Frequency-domain & error\\
\hline
\tstrut$\partial_t\psi^\text{R}$ & $10M$ & $3.750211\times 10^{-5}$
                                         & $3.750227\times 10^{-5}$ & 0.000431\% \\
\tstrut$\partial_r\psi^\text{R}$ & $10M$ & $1.380612\times 10^{-5}$
                                         & $1.378448\times 10^{-5}$ & 0.157\% \\
\hline
\tstrut$\partial_t\psi^\text{R}$ & $12M$ & $1.747278\times 10^{-5}$
                                         & $1.747254\times 10^{-5}$ & 0.00139\% \\
\tstrut$\partial_r\psi^\text{R}$ & $12M$ & $5.715982\times 10^{-6}$
                                         & $5.710205\times 10^{-6}$ & 0.101\% \\
\hline\hline
\end{tabular}
\end{table}

\begin{figure}[t]
  \centering
  \includegraphics[scale=1,angle=0]{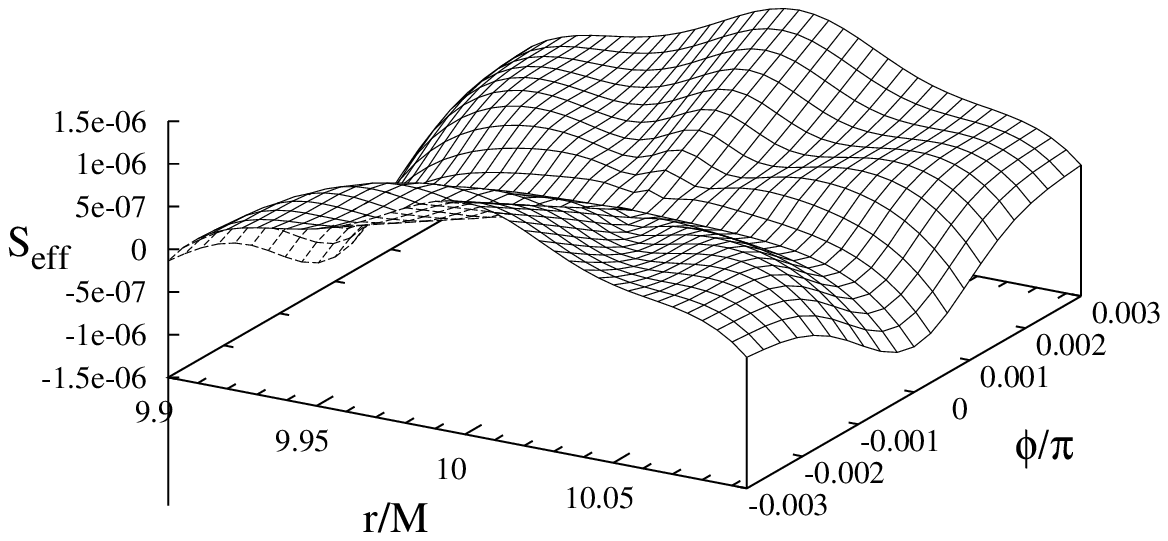}
  \caption{From \cite{VegaDet08}. The effective source
  $S_{\text{eff}}$ in the equatorial plane in the vicinity of the
  point source at $r/M=10$, $\phi/\pi=0$. Note the significant
  difference of scales with Fig.~\ref{figS}.}
  \label{figSC0}
\end{figure}

Figure \ref{figSC0} reveals the $C^0$ nature of the effective source at
the location of the particle on a dramatically different scale. It is
important to note that limited differentiability of this sort does not
introduce a small length scale into the numerical problem, and might be
treated via a special stencil in the neighborhood of the charge.

A recent collaboration with Peter Diener, Wolfgang Tichy and Ian Vega
\cite{effsource09} looks at the same test problem but involves two
distinct $3+1$ codes, which were developed completely independently. One
uses pseudo-spectral methods, the other uses a multiblock code with high
order matching across block boundaries. With a modest amount of effort
these two codes, each developed for generic numerical relativity problems,
were modified to accommodate the effective source of the scalar field and
are able to determine all components of the effective source with errors
less than $1\%$. The future of numerical $3+1$ self-force analysis looks
promising.

\section{Concluding remarks}
\label{conclusion}

Ptolemy was able to model accurately the motion of the planets in terms
of epicycles and circles about the Earth. 
However, the precise choice
of which circles and epicycles 
should be used was debated. Copernicus
realized that a much cleaner description resulted from having the
motion centered upon the sun. The two competing models were equally
able to predict the positions of the planets for the important task of
constructing horoscopes. But for understanding the laws of physics,
Newton clearly favored the Copernican model.

There appear to be two rather distinct attitudes toward calculating
the effects of the gravitational self-force for a mass $\MU$ orbiting
a black hole. Both lead to identical conclusions about physically
measurable quantities. If the motion is to be described as
accelerating in the black hole geometry, then the acceleration
depends upon the perturbative gauge choice and is not related to any
acceleration that an observer local to $\MU$ could actually measure.
If the motion is described as geodesic in the spacetime geometry
through which $\MU$ moves, then it is immediately apparent that the
only quantities worth calculating are those which are physically
measurable, or at least independent of the gauge choice. With this
second attitude, one is left with the rather satisfying 
perspective that
the effects of the gravitational self-force are neither more nor less
than the result of free-fall in a gravitational field.

In this review. I have eschewed mention of Green's functions. The
asymptotic matching perspective promoted here seems more effective to
me at getting to the physics of the gravitational self-force and less
likely to lead to mathematical confusion.

The singular field $h^{\text{S}}_{ab}$, which plays a fundamental role,
has a reasonably straightforward description in convenient locally
inertial coordinates. And it appears nearly immediately in the
DW\cite{DetWhiting03} formulation of radiation reaction via the Green's
function $G^{\text{S}}_{abc^\prime d^\prime}$ \footnote{In fact the
singular field was discovered first \cite{Det01} using matched
asymptotic expansions. And the Green's function appeared only later
during an attempt to show consistency with the usual DeWitt-Brehme
\cite{DeWittBrehme60} approach to radiation reaction.}. This Green's
function has odd acausal structure with support on the past and future
null cone of the field point and also in the spacelike related region
outside these null cones. Such causal structure is consistent with the
fact that $h^{\text{S}}_{ab}$ exerts no self-force. Based upon personal
conversations, this feature appears problematical to some. However, the
integrability condition of the perturbed Einstein equation requires
that the worldline of a point source be a geodesic. Geodesic motion is
the General Relativistic equivalent of Newtonian no motion, and the
singular field is the curved space equivalent of a Coulomb field.  Not
much is happening at the source or to the singular field. I cannot
imagine that such behavior somehow leads to an effect that might be
described as acausal.

The S-field $h^{\text{S}}_{ab}$ is defined via an expansion in a
neighborhood of the source and does not depend upon boundary conditions,
and the restriction to geodesic motion precludes any unexpected behavior
of the point mass in either the past or the future. The S-field is
precisely the nearly-Newtonian monopole field with minor tidal distortions
from the surrounding spacetime geometry.

While orbiting a black hole, Einstein's apple emits gravitational
waves and spirals inward. However, the apple is in free fall and not
accelerating. In fact, it is not moving in its locally inertial frame
of reference, and is aware of neither
 its role as the source of any radiation nor of its role acting out
the effects of radiation reaction.

S.\ Chandrasekhar was fond of describing a conversation with the
sculptor Henry Moore. In his own words, Chandra ``had the occasion to
ask Henry Moore how one should view sculptures: from afar or from near
by. Moore's response was that the greatest sculptures can be
viewed---indeed should be viewed---from all distances since new aspects
of beauty will be revealed at every scale.''\cite{MTBH} 
 The self-force analysis in General Relativity also
reveals different aspects when viewed from afar and when viewed from
near by.
 From afar a small black hole dramatically emits gravitational waves while
inspiralling toward a much larger black hole. From near by the small
hole reveals the quiet simplicity and grace of geodesic motion.
 Rather than ``beauty,'' a satisfying sense of physical consistency
is ``revealed at every scale.''

\section*{Acknowledgements}

My understanding of gravitational self-force effects has evolved over
the past decade in large part in discussions with colleagues during the
annual Capra meetings. I am deeply indebted to the organizers and
participants of these fruitful meetings. And I am particularly pleased
to have had recent collaborators Leor Barack, Peter Diener, Eric
Poisson, Norichika Sago, Wolfgang Tichy, Ian Vega, and Bernard Whiting,
who individually and as a group have kept me on track and moving
forward.

This work was supported in part by the National Science Foundation,
through grant number PHY-0555484 with the University of Florida. Some
of the numerical results described here were preformed at the
University of Florida High-Performance Computing Center (URL:
http://hpc.ufl.edu).

 \bibliographystyle{prsty}


\end{document}